\documentclass{article} 
 \pdfoutput=1 

 \usepackage{graphicx}

\usepackage{caption}
\usepackage{subcaption}

\usepackage{amssymb}

\usepackage{amsmath}
\usepackage{bm} 
\usepackage{amsfonts}

\usepackage{color}
\usepackage{mathdots}
\usepackage[pagebackref=true]{hyperref}
\usepackage{xspace} 
\usepackage{array}
\usepackage[binary-units=true,]{siunitx}
\usepackage{arydshln}
\usepackage{enumitem}
\usepackage{tabularx}
\usepackage{txfonts}
\usepackage[shortcuts]{extdash}
\usepackage{authblk}
\usepackage{geometry}

\usepackage{multirow}

\renewcommand{\arraystretch}{1} 

\newcommand{\ar}{\arraycolsep=1.5pt\def\arraystretch{1.2}}

\usepackage{booktabs}
\newcommand{\ra}[1]{\renewcommand{\arraystretch}{#1}}

\newcommand{\choptrey}{CHOPtrey\xspace} 
\newcommand{\CHOPA}{Contextual \& Hierarchical Ontology of Patterns\xspace} 
\newcommand{\CHOPR}{Computationally Hasty Online Prediction\xspace} 
\newcommand{\CHOPO}{Causal Hopping Oblivious Polynomials\xspace} 
\newcommand{\chopatt}{CHOP{\small{att}}\xspace} 
\newcommand{\chopred}{CHOP{\small{red}}\xspace} 
\newcommand{\chopoly}{CHOP{\small{oly}}\xspace} 

\newcolumntype{x}[1]{>{\centering\let\newline\\\arraybackslash\hspace{0pt}}p{#1}} 

\newcommand{\NN}{{\ensuremath{\mathbb N}}}

\newcommand{\ol}[1]{\ensuremath{\overline{{#1}}}}

\newcommand{\mk}[2]{\ensuremath{m_{{#1},{#2}}}}
\newcommand{\Sk}[2]{\ensuremath{z_{{#1},{#2}}}}
\newcommand{\Zk}[2]{\ensuremath{\bm{Z}_{{#1},{#2}}}}
\newcommand{\mmk}[3]{\ensuremath{\overline{m}_{{#1},{#2},{#3}}}}
\newcommand{\MMk}[3]{\ensuremath{\overline{\bm{m}}_{{#1},{#2},{#3}}}}
\newcommand{\SSk}[3]{\ensuremath{\overline{z}_{{#1},{#2},{#3}}}}
\newcommand{\ZZk}[3]{\ensuremath{\overline{\bm{Z}}_{{#1},{#2},{#3}}}}

\newcommand{\bigO}[1]{\ensuremath{\mathop{}\mathopen{}\mathcal{O}\mathopen{}\left(#1\right)}}

\usepackage{tikz}
\usetikzlibrary{matrix,calc,shapes}

\tikzset{
  treenode/.style = {shape=rectangle, rounded corners,
                     draw, anchor=center,
                     text width=12em, align=center,
                     top color=white, bottom color=blue!20,minimum height=4ex},
  decision/.style = {treenode, diamond, aspect=2, inner sep=-4pt},
  root/.style     = {treenode, ellipse, bottom color=red!30,text width=10em,minimum height=6ex},
  finish/.style   = {root, bottom color=green!40},
}

\def\keywords{\vspace{.5em}
{\textit{Keywords}:\,\relax%
}}

\title{\choptrey: contextual online polynomial extrapolation for enhanced multi-core co-simulation of complex systems\footnote{Published in  Simulation: Transactions of the Society for Modeling and Simulation International, 2017, \url{http://dx.doi.org/10.1177/0037549716684026}
}}

\author[1]{Abir Ben Khaled-El Feki}
\author[1,2]{Laurent Duval}
\author[3]{Cyril Faure}
\author[4]{Daniel Simon}
\author[1]{Mongi Ben Gaid}
\affil[1]{IFP Energies nouvelles, 1 et 4 avenue de Bois-Pr\'eau, 92852 Rueil-Malmaison, France}
\affil[2]{University Paris-Est, LIGM, ESIEE Paris, 93162 Noisy-le-Grand, France}
\affil[3]{CEA List, Nano-INNOV, 8 avenue de la Vauve, 91120 Palaiseau, France}
\affil[4]{INRIA and LIRMM - CAMIN team, 860 Rue Saint Priest, 34095 Montpellier Cedex 5, France}

\begin{document}

\maketitle

\begin{abstract}
The growing complexity of Cyber-Physical Systems (CPS), together with increasingly available parallelism provided by multi-core chips, fosters the parallelization of simulation. Simulation speed-ups are expected from co-simulation and parallelization based on model splitting into weak-coupled sub-models, as for instance in the framework of Functional Mockup Interface (FMI). However, slackened synchronization between sub-models and their associated solvers running in parallel introduces integration errors, which must be kept inside acceptable bounds. 

\choptrey denotes a forecasting framework enhancing the performance of complex system  co-simulation, with a trivalent articulation. 
First, we consider the framework of a \CHOPR system (\chopred). It allows to improve the trade-off between integration speed-ups, needing large communication steps, and simulation precision, needing frequent updates for model inputs. 
Second, smoothed adaptive forward prediction improves co-simulation accuracy. It is obtained by past-weighted extrapolation based on \CHOPO (\chopoly). 
And third, signal behavior is segmented to handle the discontinuities of the exchanged signals: the segmentation is performed in a \CHOPA (\chopatt).

Implementation strategies and simulation results demonstrate the framework ability to  adaptively relax data communication constraints beyond synchronization points which sensibly accelerate simulation.  
The  \choptrey framework extends the range of applications of standard Lagrange-type methods, often deemed unstable. The embedding of predictions in lag-dependent smoothing and discontinuity handling demonstrates its practical efficiency.
\end{abstract}

\keywords{
parallel simulation; Functional Mockup Interface; smoothed online prediction; causal polynomial extrapolation; context-based decision; internal combustion engine.}


\section{Introduction}
Intricacy in engineered systems increases simulator complexity. However, most existing simulation softwares are currently unable to exploit multi-core platforms, as they rely on sequential Ordinary Differential Equations (ODE) and Differential Algebraic Equations (DAE) solvers. Co-simulation approaches can provide significant improvements by allowing to jointly simulate models coming from different areas, and to validate both individual behaviors and their interactions~\cite{COSIM}. Different simulators may be exported from original authoring tools, for instance as Functional Mock-up Units (FMUs), and then imported in a co-simulation environment. Hence, they cooperate at run-time, thanks to Functional Mockup Interface (FMI) definitions~\cite{fmi:modelica2011} for their interfaces, and to the master algorithms of these environments. 

Co-simulation has shown an important potential for parallelization (see~\cite{Benkhaled_2014} for a review). Meanwhile, synchronization between the different sub-models is required due to their mutual dependencies. It avoids the propagation of numerical errors in simulation results and guarantees their correctness. Unfortunately, synchronization constraints also lead some processors into waiting periods and idle time. This decreases the potential efficiency of the threaded parallelism existing in multi-core platforms.

To overcome this limitation and exploit the available parallelism more efficiently, the dependencies constraints between parallel sub-models should be relaxed as far as possible while preserving an acceptable accuracy of the simulation results. This can be performed by a well-grounded system decomposition, tailored to data dependency reduction between the sub-models. For instance, the method proposed in~\cite{sjolund:2010} for distributed simulation uses transmission line modeling (TLM) based on bilateral delay lines: decoupling points are chosen when variables change slowly and the time-step of the solver is relatively small.

Unfortunately, perfect decoupling cannot often be reached and data dependencies still exist between the different blocks. Thus, tight synchronization is required between blocks using small communication steps. This greatly limits the possibilities to accelerate the simulation and eventually reach  the real-time. 

We propose in this work a \CHOPR framework (\chopred) to stretch out synchronization steps with negligible precision changes in the simulation, at low-complexity.
It is based on a \CHOPA (\chopatt) that selects appropriate \CHOPO (\chopoly) for signal forecasting, allowing data exchange between sub-models beyond  synchronization points.

This paper is organized as follows. We first review challenges as well as related work and summarize contributions in Section~\ref{sec_contributions}. 
Section~\ref{sec:Pb_form_mot} presents a formal model of a hybrid dynamical system and the motivations for twined parallel simulation and extrapolation.
The background on prediction and the proposed \CHOPO (\chopoly) are developed in Section~\ref{sec_causal-pol-pred}, with details in \ref{sec:appendix}. 
Then, the principles of the \CHOPA (\chopatt) for the management of hybrid models are exposed in Section~\ref{sec:context}. Finally, the methodology's performance is assessed in Section~\ref{sec:results} using an internal combustion engine model described in Section~\ref{sec:CS}.

\section{Related work and contributions\label{sec_contributions}}
The continuous systems usually attain high integration speeds with variable-step solvers. The major challenge for hybrid systems resides in their numerous discontinuities that prevent similar performance.
Since we are especially interested in modular co-simulation~\cite{Benkhaled_2014}, it is shown in~\cite{Benkhaled_A_2012_ECOSM} that integrating each sub-system with its own solver allows to avoid interrupts coming from unrelated events. Moreover event detection and location inside a sub-system can be processed faster because they involve a smaller set of variables. However, partitioning the original complex model into several lesser complex models may add virtual algebraic loops, therefore involving delayed outputs, even with an efficient execution order. 

To take advantage of the model splitting without adding useless delays, we propose in~\cite{Benkhaled_A_Simpra} a new co-simulation method based on a refined scheduling approach. This technique, denoted ``RCosim'', retains the speed-up advantage of modular co-simulation thanks to the parallel execution of the system's components. Besides, it improves the accuracy of the simulation results through an offline scheduling of operations that takes care of model input/output dynamics. 

However, in practical applications, current co-simulation set-ups use a constant communication grid shared by all the models. In fact, the size of communication steps has a direct impact on simulation errors, and effective communication step control should rely on online estimations of the errors induced by slackened exchange rates. 
Schierz \emph{et al.}~\cite{schierz2012co} propose the use of adaptive communication step sizes to better handle the various changes in model dynamics. Meanwhile, the stability of multi-rate simulators with adaptive steps needs to be carefully assessed, for example based on errors propagation inside modular co-simulations~\cite{Arnold_M_2010_j-comput-nonlinear-dynam_stability_smtimcmsm}. 

Data extrapolation over steps is also expected to enhance the simulation precision over large communication steps. Nevertheless, actual complex systems (CPS) usually present non-linearities and discontinuities, making hard to predict their future behavior from past observations only. Moreover, the considered models are generated using the Simulink Coder target or the FMI for Model Exchange framework, which does not provide input derivatives (in contrast with the FMI for Co-Simulation architecture). Hence polynomial prediction cannot always correctly extrapolate. For example, \cite{Arnold_M_2008_incoll_numerical_msad} uses a constant, linear or quadratic extrapolation and a linear interpolation to improve the accuracy of the modular time integration. This method is successful for non-stiff systems but fails in the stiff case.

In the purpose of defining a method for the parallel simulation of hybrid dynamical systems, our previous work on a single-level context-based extrapolation~\cite{BenKhaled_A_2014_p-modelica_context-based_pessfmcsufmi} accounts for steps, stiffness, discontinuities or weird behaviors. It uses adapted extrapolation to limit excessively wrong prediction. It shows that properly-chosen context-based extrapolation, combined with model splitting and parallel integration, can potentially improve the speed vs. precision trade-off needed to reach real-time simulation.

In~\cite{BenKhaled_A_2014_p-modelica_context-based_pessfmcsufmi}, context-based extrapolation is exclusively intended  for FMU models and extrapolation is performed  on integration steps only. This paper improves upon the preceding results with a better-grounded methodology, by: 

\begin{itemize}
	\item adding oblivion to past samples through a weighting factor in polynomial prediction;
	\item increasing the computational efficiency via a novel matrix formulation;
	\item adding an online error evaluation at each communication step to select the best context and weighting factor at the forthcoming step, further minimizing extrapolation errors;
	\item adding a hierarchy of decisional  and functional contexts to improve prediction strategy.
\end{itemize}

\section{Problem formalization and motivation\label{sec:Pb_form_mot}}
\subsection{Model definition\label{sec:form}}
Complex physical systems are generally modeled by hybrid non-linear ODEs or DAEs. The hybrid behavior is due to the discontinuities, raised by events triggered off by the crossing of a given threshold (zero-crossing). It plays a key role in the simulation complexity and speed. Indeed, more events slow down numerical integration. 

Let us provide a formal model, considering a hybrid dynamic system $\Sigma$ whose continuous state evolution is governed by a set of non-linear differential equations:
\[
\begin{array}{ccccc}
\dot{\bm{X}} & = & \bm{f}(t,\bm{X},\bm{D},\bm{U}_{\operatorname{ext}})& \operatorname{for}& t_n \leq t < t_{n+1}\,,\\
\bm{Y}_{\operatorname{ext}} & = & \bm{g}(t,\bm{X},\bm{D},\bm{U}_{\operatorname{ext}})\,,&&
\end{array}
\]
where $\bm{X} \in \mathbb{R}^{n_{X}}$ is the continuous state vector, $\bm{D} \in \mathbb{R}^{n_{D}}$ is the discrete state vector,
$\bm{U}_{\operatorname{ext}} \in \mathbb{R}^{n_{U_{\operatorname{ext}}}}$ is the external input vector, $\bm{Y}_{\operatorname{ext}} \in \mathbb{R}^{n_{Y}}$ is the external output vector and $t \in \mathbb{R}^+$ denotes the time.
The sequence $(t_n)_{n \geq 0}$ of strictly increasing time instants represents discontinuity points called \emph{state events}, which are the roots of the equation:
\[
\bm{h}(t,\bm{X},\bm{D},\bm{U}_{\operatorname{ext}}) = 0\,.
\]
The function $\bm{h}$ is usually called \emph{zero-crossing} function or \emph{event indicator}, used for \emph{event detection} and \emph{location}~\cite{mosterman:ifac08}.
At each time instant $t_n$, a new continuous state vector can be computed as a result of the \emph{event handling}:
\[
\bm{X}(t_n) = \bm{I}(t_n,\bm{X},\bm{D},\bm{U}_{\operatorname{ext}})\,,
\]
and a new discrete state vector can be computed as a result of a discrete state update:
\[
\bm{D}(t_n) = \bm{J}(t_{n-1},\bm{X},\bm{D},\bm{U}_{\operatorname{ext}})\,.
\]
If no discontinuity affects a component of $\bm{X}(t_n)$, the right limit of this component will be equal to its value at $t_n$. 
This hybrid system model is adopted by several modeling and simulation environments and is underlying the FMI specification~\cite{fmi:modelica2011}.
\par
We assume that $\Sigma$ is well-posed, in the sense that a unique solution exists for each admissible initial conditions $\bm{X}(t_0)$ and $\bm{D}(t_0)$ and that consequently $\bm{X}$, $\bm{D}$, $\bm{U}_{\operatorname{ext}}$, and $\bm{Y}_{\operatorname{ext}}$ are piece-wise continuous functions, i.e. continuous on each sub-interval $]t_n, t_{n+1}[$. 

\subsection{Model parallelization with modular co-simulation\label{sec:mot_par}}
\begin{figure} [!htb]
\begin{center}
\includegraphics[width=0.75\columnwidth]{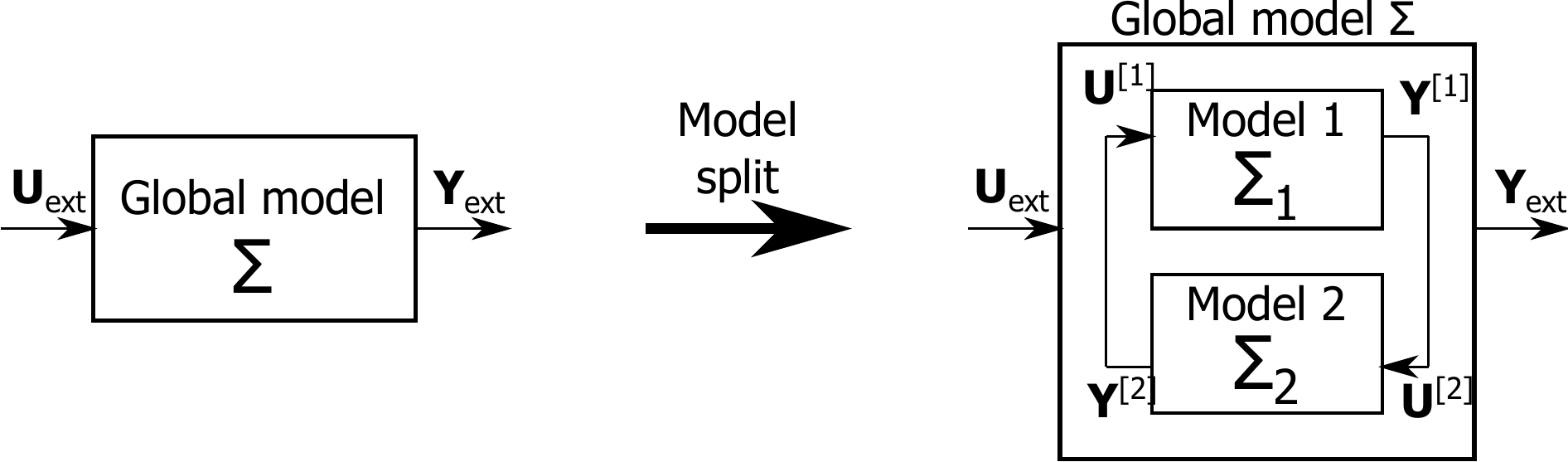} 
\caption{System splitting for parallelization.} 
\label{fig:sys2}
\end{center}
\end{figure}
To execute the system in parallel, the model must be split into several sub-models. For simplicity, assume that the system is decomposed into two separate blocks denoted Model~1 and Model~2, in Figure~\ref{fig:sys2} with $\bm{X}=[\bm{X}^{[1]}\;\bm{X}^{[2]}]^T$ and $\bm{D}=[\bm{D}^{[1]}\;\bm{D}^{[2]}]^T$, where $^T$ denotes the matrix transpose. 
Therefore, the sub-systems can be written as: 
\[
\begin{array}{l}
\left\{
  \begin{array}{lll}
   \bm{\dot X}^{[1]} &\hspace{-3mm}=&\hspace{-3mm}\bm{f}^{[1]}(t,\bm{X}^{[1]},\bm{D}^{[1]},\bm{U}^{[1]},\bm{U}_{\operatorname{ext}})\,, \\
   \bm{Y}^{[1]}&\hspace{-3mm}=&\hspace{-3mm}\bm{g}^{[1]}(t,\bm{X}^{[1]},\bm{D}^{[1]},\bm{U}^{[1]},\bm{U}_{\operatorname{ext}})\,, \\
  \end{array}
 \right.
\\\\
 \left\{
  \begin{array}{lll}
   \bm{\dot X}^{[2]}&\hspace{-3mm}=&\hspace{-3mm}\bm{f}^{[2]}(t, \bm{X}^{[2]},\bm{D}^{[2]},\bm{U}^{[2]},\bm{U}_{\operatorname{ext}})\,, \\
   \bm{Y}^{[2]}&\hspace{-3mm}=&\hspace{-3mm}\bm{g}^{[2]}(t, \bm{X}^{[2]},\bm{D}^{[2]},\bm{U}^{[2]},\bm{U}_{\operatorname{ext}})\,. \\
  \end{array}
 \right.
 \end{array}
\]
Here, $\bm{U}^{[1]}$ are the inputs needed for Model~1 ($\Sigma_1$), directly provided by the outputs $\bm{Y}^{[2]}$ produced by Model~2 ($\Sigma_2$).
Similarly, $\bm{U}^{[2]}$ are the inputs needed for Model~2 directly provided by the outputs $\bm{Y}^{[1]}$ produced by Model~1.
Our approach generalizes to any decomposition of the system $\Sigma$ into $B$ blocks, where each block is indexed by $b\in\{1,\ldots,B\}$.
\subsection{Model of computation}
\begin{figure}[!htb]
   \centering
   \includegraphics[width=0.75\columnwidth]{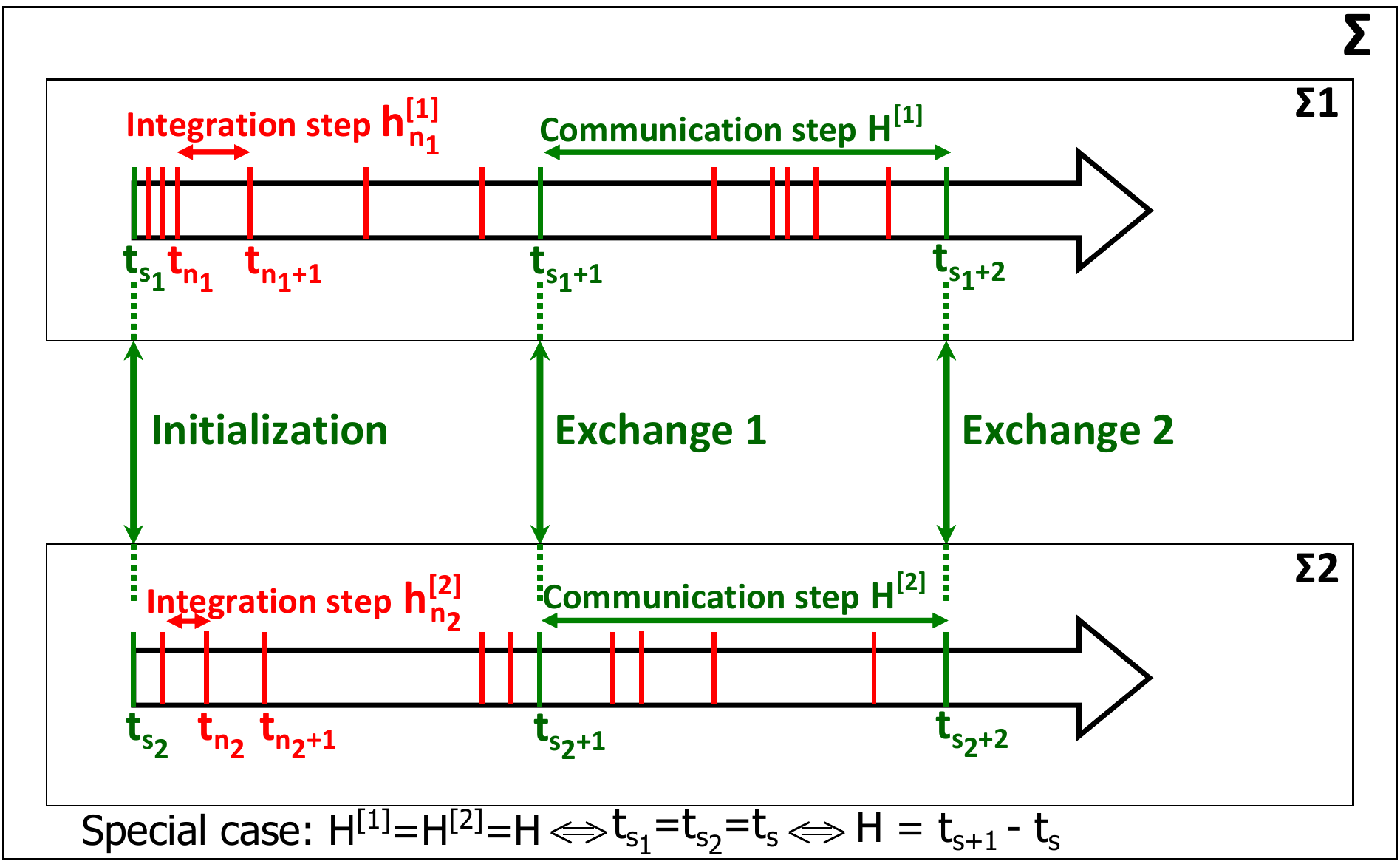}
   \caption{$\Sigma$ split into $\Sigma_1$ and $\Sigma_2$ for parallel simulation.}
   \label{fig:sys}
  \end{figure}

To perform the numerical integration of the whole multi-variable system, each of these simulators needs to exchange, at communication (or synchronization) points  $t_{s_b}$ ($b=1$ or $b=2$), the data needed by the other (see Figure~\ref{fig:sys}).
To speed up integration, the parallel branches must be as independent as possible, so that they are synchronized at a rate $H^{[b]}=t_{s_b+1}-t_{s_b}$, by far slower than their internal integration step $h^{[b]}_{n_b}$ ($H^{[b]}\gg h^{[b]}_{n_b}$). Therefore, between communication points, each simulator integrates at its own rate (assuming a variable-step solver), and considers that data incoming from others simulators is held constant.
	
\begin{figure}[!htb]
   \centering
   \includegraphics[width=0.7\columnwidth]{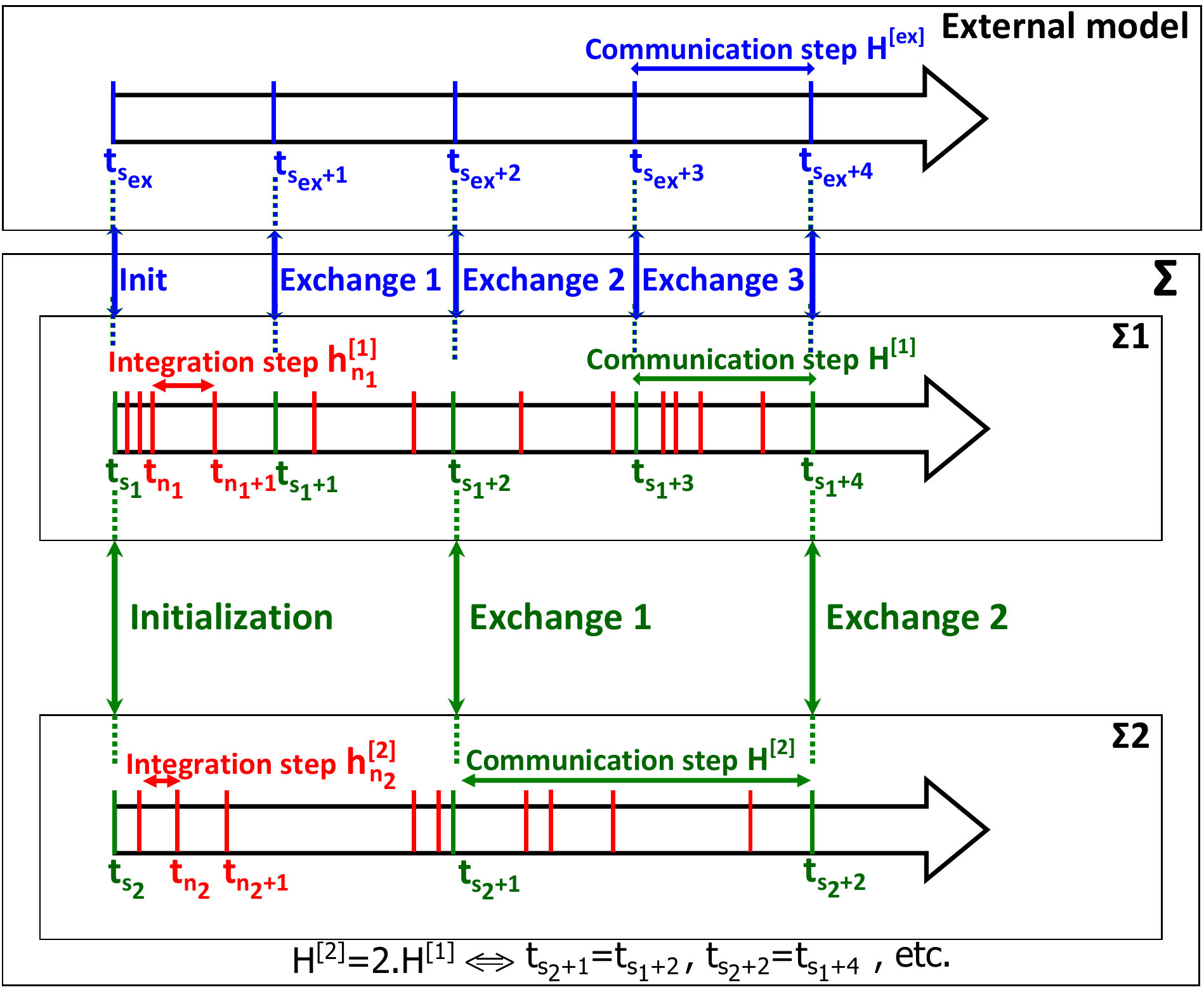}
   \caption{Parallel multi-rate simulation.}
   \label{fig:sys4}
  \end{figure}

Besides, when $H^{[b_1]}\neq H^{[b_2]}$, data incoming from other slower simulators are held constant.
For instance in Figure~\ref{fig:sys4}, Model~1 needs to communicate with an external model two times faster than Model~2). Data incoming from Model~2 is held constant during $2H^{[1]}$, potentially causing aliasing.
  
It is likely that large and multi-rate communication intervals allow to speed up the numerical integration, but may result in integration errors and poor confidence in the final result. 
\subsection{Motivation for extrapolation\label{sec:mot_extra}}
Modeling the errors induced by slackened synchronization is a first direction in improving the trade-off between integration speed and accuracy.
\subsubsection{Integration errors and parallelism}
Error evaluation and convergence analysis were performed in~\cite[Chapter 9]{Benkhaled_2014} for different schemes (sequential and parallel modular co-simulation, models with real and artificial loops, mono-rate and multi-rate co-simulation, etc.). Assuming that the $b^{\operatorname{th}}$ sub-model is connected to \mbox{$B_c-1$} other sub-models, its global error on states \eqref{eq:ineqX} and outputs~\eqref{eq:ineqY} are bounded in Landau-Bachmann $\bigO{}$ notation. Errors are functions is function of the communication step (or synchronization interval) $H=\max H^{[b]}$, $b\in\left\{1,\ldots, B_c\right\}$, the integration time-step $h$ and the order of accuracy $p$ of the used numerical solver \cite[Chapter 4]{Benkhaled_2014}:
\begin{eqnarray} 
\left\|\bm{X}^{[b]}(t_{n+1}) - \bm{X}^{[b]}_{n+1}\right\| &\leq& \bigO{h^{p}}+\bigO{H}\,,\label{eq:ineqX}\\
\left\|\bm{Y}^{[b]}(t_{n}) - \bm{Y}^{[b]}_{n}\right\| &\leq& \bigO{h^{p}}+\bigO{H}\,.\label{eq:ineqY}
\end{eqnarray}

Both global errors are clearly bounded by two terms. The first term is related to the applied numerical solver, more specifically the time-step and the order. The worst case scenario on the bounding would be the maximum used integration step $h$ and order $p$. The second term is related to the size of the communication step $H$. However, the weight of each term on the error is deeply related to the size of the communication step relatively to the integration step. Based on the same approach as in~\cite{Arnold_M_2008_incoll_numerical_msad}, it is clear that the communication step $H$ dominates the error when $H \gg h$.
	
Therefore, considering a split model and a parallel execution, a trade-off must be found between acceptable simulation errors, thanks to tight enough synchronization, and simulation speed-up thanks to decoupling between sub-models.
\subsubsection{Contribution of the extrapolation\label{sec_contribution-extrapolation}}
To add a degree of freedom to this trade-off achievement, we propose to extrapolate model inputs to compensate the stretching out of communication steps between sub-models. In fact, it was proven previously that numerical solutions are first order accurate, $\bigO{H}$, when choosing larger communication steps $H$. Holding inputs constant between two synchronization intervals plays the role of a zeroth-order hold (ZOH, constant extrapolation). To generalize the error bound, the $\bigO{H}$ term can then be replaced with \bigO{H^{\delta+1}}, where $\delta$ is the extrapolation   degree. Using for example linear ($\delta=1$) or quadratic ($\delta=2$, \ref{sec:prediction-parabola}) extrapolation instead of constant update ($\delta=0$) has the potential to  reduce the bound of simulation errors, as proven in~\cite[p. 238 \emph{sq.}]{Arnold_M_2008_incoll_numerical_msad}.
However, the difficulty with extrapolation is that it could be sensitive for different reasons:
\begin{itemize}
  \item there exists no universal prediction scheme, efficient for every signal;
	\item prediction should be efficient: causal, sufficiently fast and reliable;
	\item standard polynomial prediction may fail in stiff cases~\cite{Arnold_M_2008_book_simulation_tad} (cf. Section~\ref{sec:context} for details).
\end{itemize}
\par
We choose to base our extrapolation on polynomial prediction, which allows fast and causal calculations. In this situation, the rationale is that the computing cost of a low-order polynomial predictor would be by far smaller than the extra model computations needed by shorter communication steps. Since such predictions would be accurate neither for any signal (for instance, blocky versus smooth signals) nor for any signal behavior (slow variations versus steep onsets), we borrow the context-based approach from lossless image encoders~\cite{Weinberger_M_2000_j-ieee-tip_loco1_licapsjpegls}, such as GIF (Graphics interchange format)  or PNG (Portable Network Graphics) compression formats. The general aim of these image coders is to predict a pixel value based on a pattern of causal neighboring pixels. Compression is achieved when the prediction residues possess smaller intensity values, and more generally a sparser distribution (concentrated around close-to-zero values) than that of pixels in the original image. They may therefore be coded on smaller ``bytes'', using entropy coding techniques. In images, one distinguishes basic ``objects'' (smooth-intensity varying regions, edges with different orientations). Based on simple calculations over prediction patterns, different behaviors are inferred (e.g. flat, smooth, $+45^o$ or $-45^o$ edges, etc.). Look-up table predictors are then used, depending on the context.
\par
In the proposed approach, we build a heuristic table of contexts (Section~\ref{sec:context}) based on a short frame of past samples, and affect pre-computed polynomial predictors to obtain  context-dependent extrapolated values. We now review the principles of extrapolation.

\section{\chopoly: \CHOPO for low-complexity extrapolation\label{sec_causal-pol-pred}}
\subsection{Background on prediction, real-time constraints and extrapolation strategies\label{smallpox_background}}

The \choptrey framework requires a dedicated instance of forecasting for discrete time series. The neighboring topics of prediction, interpolation  or extrapolation represent a large body of knowledge in signal processing~\cite{Wiener_N_1949_book_extrapolation_issts,Meijering_E_2002_j-proc-ieee_chr_iaamsip}, econometrics~\cite{Brown_R_1962_book_smoothing_fpdts}, control~\cite{Box_G_2008_book_time_safc,Valiviita_S_1999_j-ieee-tie_polynomial_pfcir} and numerical analysis \cite{Richardson_L_1911_j-phil-trans-r-soc-a_approximate_asfdppideasmd}. They are paramount in complex systems that live, sample  and communicate at different rates. Their use in simulation might be milder. There exists a natural and common belief that signal extrapolation may introduce additional error terms and cause numerical instability \cite{Arnold_M_2013_j-arch-mech-eng_error_aeecsfmimecsv20,Stettinger_G_2014_p-ieee-cdc_model-based_apbids}. Building upon our previous work~\cite{BenKhaled_A_2014_p-modelica_context-based_pessfmcsufmi}, two additions diminish the importance of extrapolation caveats: past sample weighting (oblivious polynomials) and a hierarchy of contexts defined by a pattern ontology.

As we operate under real-time conditions, implying strong causality,  only a tiny fraction of time series extrapolation methods are practically applicable.
For instance, the theory of multi-rate filter banks bears formal similarities with distributed simulation or co-simulation, and have long been deployed for interpolation, extrapolation, and reduction of computational needs~\cite{Bellanger_M_1974_j-ieee-tassp_interpolation_ercsdf}. Such systems allow optimized noise handling at variable redundancy rates~\cite{Gauthier_J_2009_j-ieee-tsp_optimization_socfb}. 
The overall propagation delay is harmless in traditional signal and image processing (for compression or signal restoration). However, it prevents their use in distributed simulation. Additionally, the decomposition of signals into frequency bands is not adapted to simulation data, that sometimes exhibit stiff behavior, preventing the use of band-limited extrapolation \cite{Liu_V_1989_p-iscas_finite_lbleds}. We alternatively  propose to feed a bank of weighted smoothing polynomials, whose outputs are selected upon both simulation and real-time behaviors, organized in a \CHOPA (\chopatt, Section\ref{sec:context}).
 
ZOH (zeroth-order hold) or nearest-neighbor extrapolation is probably the most natural, the less hypothetical, and the less computationally expensive forecasting method. It consists in using the latest known sample as the predicted value. It possesses small (cumulative) errors when the time series is relatively flat  or when its sampling rate is sufficiently high, with respect to the signal dynamics. In other words, it is efficient when the time series is sampled fast enough to ensure small variations between two consecutive sampling times. 
However, it indirectly leads to under-sampling or aliasing related disturbances \cite{Benedikt_M_2013_j-math-comp-model-dyn-syst_modelling_anicpcs}. They affect the signal information content, including its derivatives, and consequently its processing, for instance through differential equation systems. They appear as quantization-like noise~\cite[p. 609 sq.]{Hamming_R_1973_book_num_mse}, delay induction, offset or bump flattening. As a remedy, higher order extrapolation  schemes can be used. They consist in fitting a $\delta$-degree polynomial through $\delta+1$ supporting points \cite[p. 15 sq.]{Friedrich_M_2011_phd_parallel_csms}. The most commons are the predictive first-order \cite{Arnold_M_2007_incoll_multi-rate_tilsmsm,Hoher_S_2011_j-simpat_contribution_rtscfmmtnc} or the second-order hold  (FOH or SOH) \cite{Benedikt_M_2013_j-math-comp-model-dyn-syst_modelling_anicpcs}. They result in linear or parabolic extrapolation.
Polynomial parameters are  obtained in a standard way with Lagrange polynomials. Hermite polynomials can be used when one is interested in extrapolating derivatives as well, and their relative stability  has been studied \cite[p. 61 sq.]{Busch_M_2012_phd_effizienten_ks}. Both, although being exact at synchronization  points, tend to oscillate amidst  them. To avoid introducing discontinuities, extrapolation can be smoothed with splines \cite{Oh_S_2016_j-energies_co-simulation_fpa} or additional interpolation \cite{Busch_M_2016_j-zamm_continuous_atcsmansle}.

In our co-simulation framework, communication intervals are not chosen arbitrarily small for computational efficiency. 
Thus, the slow variation of inputs and outputs cannot be ensured in practice. Provided a cost vs. error trade-off is met, borrowing additional samples from past data and using higher-order extrapolation schemes could be beneficial. Different forecasting methods of various accuracy and complexity may be efficiently evaluated. We focus here on online extrapolation with causal polynomials, for  simplicity and ease of implementation, following initial work in \cite[Chap. 16]{Faure_C_2011_phd_real-time_spmthlv}. Their main feature is that they are obtained in a least-squares fashion: we do not impose that they pass  exactly through supporting points. However, they do so when  the degree $\delta$ and the number of supporting points $\delta+1$ is set as above.
We improve on~\cite{BenKhaled_A_2014_p-modelica_context-based_pessfmcsufmi} by  adding a time-depending oblivion faculty to  polynomial extrapolation  with an independent power weighting on past samples. It  accounts for memory depth changes required to adapt to sudden variations. Computations are performed on  frames of samples hopping at synchronization steps, allowing a low-complexity matrix formulation. The resulting \CHOPO (\chopoly) are described next,   evaluated on the case study described in Section \ref{sec:CS}  and tested in Section~\ref{sec:results}.
\subsection{Notations\label{smallpox_notations}}
The first convention abstracts a sampling framework independent of the actual sampling period $H$ and the running time index. This amounts to considering  a unit communication step for any signal $u(t)$, i.e. $H=1$, and to using a zero-reindexing convention: the last available sample is indexed by $0$ ($u_0$), and the previous samples are thus indexed backwards: $u_{-1}$, $u_{-2}$, $u_{-3}$,\ldots This simplification possesses two main traits:
\begin{enumerate}
\item local indices hop at each communication step and become independent of the actual timing and communication rate;
\item some intermediate computations may thus be performed only once, reducing the risk of cumulative numerical errors~\cite{Beckermann_B_1999_incoll_sensitivity_lspa}.
\end{enumerate}

We can either use a recursive formulation with infinite memory, or a finite prediction frame. The first option is used for instance in adaptive filtering or Kalman estimation. It generally includes oblivion, implemented for instance using  a scalar forgetting factor $w_l$, $l\in ]-\infty,\ldots,-1,0]$, assigning a decreasing  weight, often exponential~\cite{Gardner_E_2006_j-int-j-forecast_exponential_ssap2}, to older error samples.
We instead consider the second option with a finite frame of the $\lambda$ last consecutive past samples $\{u_{1-\lambda},u_{2-\lambda},\ldots,u_{0}\}$. Limited-sized buffers are more natural for polynomial forecasting, and they also appear beneficial to reset computations in the case of sudden changes in the data, as discussed in Section~\ref{subsec:context}. To emulate a variable memory depth weight, we choose a piece-wise power weighting with order $\omega\ge 0$. It can be expressed as follows, without  $\lambda$ and $\omega$ indices to lighten notations:
\begin{align}
	w_l = 
	\begin{cases}
	0	&	\operatorname{if } l < 1-\lambda\,,\\
	\left(\dfrac{\lambda+l}{\lambda} \right)^\omega	&	  1-\lambda \leq l \leq 0\,.
	\label{eq_weighting}
	\end{cases}
\end{align}

\begin{figure}[!htb]
\centering
\includegraphics[width=0.9\columnwidth]{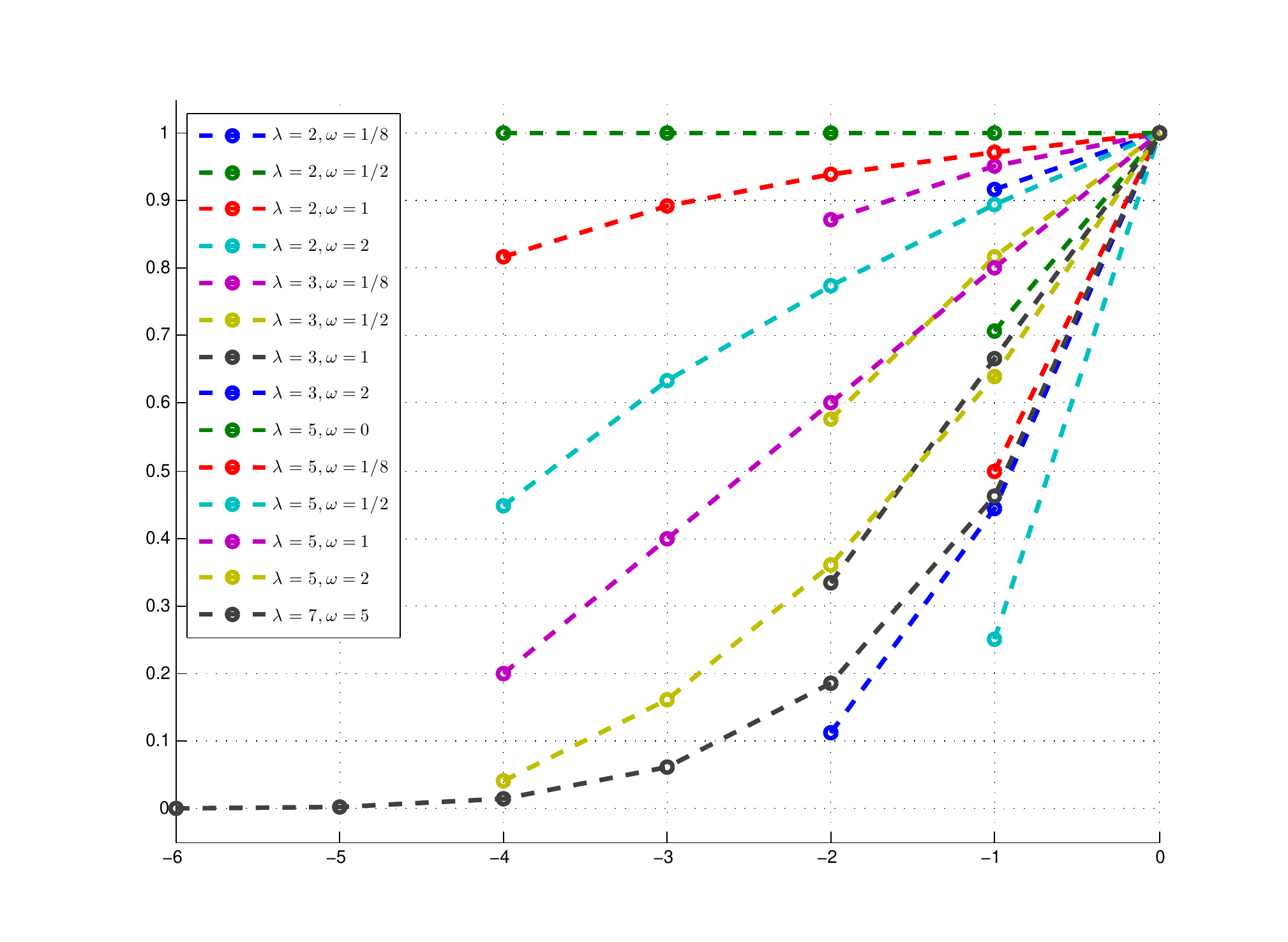}
\caption{Effect of the  weighting power $\omega$ with  different choices of $\lambda$ on the memory depth. \label{fig:weighting-factor}}
\end{figure}

Their behavior for different  powers $\omega$  is illustrated in Figure~\ref{fig:weighting-factor}. For $\omega=0$, all  samples in the frame are assigned the same importance. The higher the power, the smaller the influence of older samples.
\subsection{\chopoly: Type I, symbolic and Type II implementations\label{sec:prediction-general}}
Although the forthcoming derivations are elementary, they are rarely exposed with their complexity evaluated.
We consider polynomial predictors, performed in the least-squares sense~\cite[p. 227 sq.]{Hamming_R_1973_book_num_mse}. We denote by $P_{\delta,\lambda,\omega}$ the least-squares polynomial predictor of degree $\delta \in \mathbb{N}$, frame length $\lambda\in \mathbb{N}^*$ and weighting factor $\omega$. Its stable estimation requires that $\lambda > \delta$. The polynomial $P_{\delta,\lambda,\omega}$ is defined by the vector of length $\delta+1$, with   coefficients $\bm{a}_\delta$; hence $u(t) = a_{\delta} + a_{\delta-1}t+\cdots+a_{0}t^\delta$.
A causal prediction consists in the  estimation of unknown data at a future and relative time $\tau \ge 0$. 
Generally $\tau\in ]0,1[$, i.e., inside the time interval between the last known sample $u_0$ and the forthcoming communication step. 
The predicted value at  time $\tau$ is loosely denoted by $u_{P_{\delta,\lambda,\omega}}(\tau)$, or simply  $u(\tau)$ when the context is self-explanatory.  We refer to \ref{sec:prediction-parabola} for an introductory example on constant weighting, degree-two or parabolic prediction ($P_{2,\lambda,0}$).

In the more generic context, we  minimize the least-squares prediction error:
\begin{equation} e(\bm{a}_\delta) = 
\sum_{l=1-\lambda}^{0} \left(\dfrac{\lambda+l}{\lambda} \right)^\omega \times \left(  u_{l} - \left(\sum_{d=0}^\delta a_d l^{\delta-d}\right) \right)^2 \,.\label{eq:error-d-l-w}\end{equation}
If one chooses $\omega = 0$, and $\lambda = \delta+1$, one recovers  the Lagrange polynomials, and the error $e(\bm{a}_\delta)$ is exactly zero. Consequently, \chopoly encompasses standard extrapolation used in co-simulation (\emph{e.g.} ZOH, FOH, SOH or higher-order). Choosing $\lambda > \delta+1$ induces  a form of smoothing, that reduces oscillations. Choosing $\omega>0$ generally promotes a better extrapolation near  the last know sample, making smoothing lag-dependent.
Both parameters take into account the possibility that the signal values at communication steps could be imperfect, due to jitter in the actual sampling or round-off errors for instance.

As $\lambda^\omega$ is a non-null constant, the derivation of \eqref{eq:error-d-l-w} with respect to $a_i$ yields $\delta+1$ equations, for each $i \in \{0,\ldots,\delta\}$:
$$\sum^{0}_{l=1-\lambda} \left(\lambda+l \right)^\omega  l^{\delta - i} u_{l} = 
\sum^{\delta}_{d=0} \left[ \sum^{0}_{l=1-\lambda}  \left(\lambda+l \right)^\omega  l^{2\delta  -d- i}  \right] a_{d}\,. $$
We define the sums of powers $\Sk{d}{\lambda} = \sum_{l=0}^{\lambda-1} {l}^d$ (see \ref{sec:prediction-parabola})  and the  weighted sums of powers  $\SSk{d}{\lambda}{\omega} = \sum_{l=0}^{\lambda-1} \left(\lambda-l \right)^\omega  l^{d}$. They can be rewritten, if $\omega \in \NN$:
$$\SSk{d}{\lambda}{\omega} =  \sum_{o=0}^{\omega} (-1)^o \binom{o}{\omega} \lambda^{\omega-o} \Sk{o+d}{\lambda}\,, $$
where $\binom{o}{\omega}$ denotes the binomial coefficient.
The associated weighted  moments write, accordingly,  $\mmk{d}{\lambda}{\omega}   = \sum_{l=0}^{\lambda-1} \left(\lambda-l \right)^\omega {l }^{d}  u_{-l}$.  We refer to \ref{sec:prediction-parabola} for additional information on power sums $z_{d,\lambda} = \sum_{l=0}^{\lambda-1}l^{d}$ and moments.
The generic extrapolation pattern takes the first following matrix form (Type I):
\begin{equation}
u(\tau)  = \left[ \begin{array}{cccc}
1 & \tau &  \cdots & \tau^{\delta} 
\end{array}\right]   
\left[ \begin{array}{cccc}
  \SSk{0}{\lambda}{\omega} & -\SSk{1}{\lambda}{\omega} &  \cdots & (-1)^{\delta}\SSk{\delta}{\lambda}{\omega} \\
  -\SSk{1}{\lambda}{\omega} & \iddots & \iddots & \vdots \\
\vdots & \iddots & \iddots &  \vdots\\
  (-1)^{\delta}\SSk{\delta}{\lambda}{\omega}  &  \cdots &  \cdots    & \SSk{2\delta}{\lambda}{\omega} 
  \end{array}\right]^{-1} 
	   \left[ \begin{array}{c}  \mmk{0}{\lambda}{\omega} \\-\mmk{1}{\lambda}{\omega} \\ \vdots \\ (-1)^{\delta} \mmk{\delta}{\lambda}{\omega}\end{array}\right] \,.	\label{eq_z_hankel}
\end{equation}
We  note  $\bm{\tau}_\delta = [1, \tau,  \cdots, \tau^{\delta}]^T$ the vector of $\tau$ powers, and $$\MMk{\delta}{\lambda}{\omega} = \left[ \mmk{0}{\lambda}{\omega},-\mmk{1}{\lambda}{\omega},\cdots, (-1)^{\delta} \mmk{\delta}{\lambda}{\omega}\right]^T.$$ The inverse of the $(\delta+1) \times (\delta+1)$ Hankel matrix $\ZZk{\delta}{\lambda}{\omega}$ in \eqref{eq_z_hankel}:
$$
\left[ \begin{array}{cccc}
  \SSk{0}{\lambda}{\omega} & -\SSk{1}{\lambda}{\omega} &  \cdots & (-1)^{\delta}\SSk{\delta}{\lambda}{\omega} \\
  -\SSk{1}{\lambda}{\omega} & \iddots & \iddots & \vdots \\
\vdots & \iddots & \iddots &  \vdots\\
  (-1)^{\delta}\SSk{\delta}{\lambda}{\omega}  &  \cdots &  \cdots    & \SSk{2\delta}{\lambda}{\omega} 
  \end{array}\right]
$$
is denoted by \ZZk{-\delta}{\lambda}{\omega}. Hence, the Type-I formula for \chopoly  writes:
\begin{equation} u(\tau) = \bm{\tau}_\delta^T  \ZZk{-\delta}{\lambda}{\omega} \MMk{\delta}{\lambda}{\omega}\,. \label{eq_typeI}\end{equation}

The generic matrix formulation in \eqref{eq_typeI} is compact. 
The  matrix \ZZk{-\delta}{\lambda}{\omega} may be  computed beforehand, as a look-up table. A polynomial of degree $\delta$ involves terms of degree $2\delta$. We note that this could lead to huge values computed from large sample indices for long simulation signals. This situation is avoided by the frame-hopping zero-reindexing convention, which thus limits round-off errors and subsequent issues in the  inversion of matrices that could  become defective \cite{Kaltofen_E_2011_p-snc_fast_ehmcnnsi}.

The moment-based formulation conceals more direct symbolic formulae, linear in past samples  and polynomial in $\tau$. Some examples are given in~\ref{sec:prediction-linear}. They can be implemented with Ruffini-Horner's rule~\cite{Pena_J_2000_j-siam-j-numer-anal_multivariate_hs} for higher extrapolation orders. However, symbolic polynomial evaluations are not always  handy to implement. 

We note $\bm{u}_{\lambda} = [u_{0}, u_{-1} ,\ldots,u_{2-\lambda},u_{1-\lambda}]^T$. Then \eqref{eq_typeI} can be rewritten in a Type II form. For instance, since $\bm{\tau}_{0}^T = 1$, extrapolation with polynomial  $P_{0,\lambda,1}$ \eqref{eq_0_l_1} rewrites  as a  matrix product with $\bm{\Pi}_{0,\lambda,1} $:
\begin{equation}
u(\tau)
=\frac{2}{\lambda (\lambda+1)}
\left[ \begin{array}{cccc}
\lambda & \lambda-1 & \cdots & 1
  \end{array}\right] 
	\bm{u}_{\lambda}= \bm{\tau}_{0}^T\bm{\Pi}_{0,\lambda,1} \bm{u}_{\lambda}\,,
\end{equation}
The general Type-II formula for predictor matrices  $\bm{\Pi}_{\delta,\lambda,\omega}$ of size $(\delta+1)\times \lambda$ implements \eqref{eq_typeI} as:
\begin{equation}
u(\tau)= \bm{\tau}_{\delta}^T\bm{\Pi}_{\delta,\lambda,\omega} \bm{u}_{\lambda}\,.
	\label{eq_typeII}
\end{equation}

Such a formulation allows an efficient storage of  matrices $\bm{\Pi}_{\lambda,\delta,\omega}$ with different weighting factors. Examples are provided in Table \ref{tab_predictors_diff_omega} in \ref{sec:prediction-typeII-examples}.
Estimates of the generic number of elementary  operations required for each extrapolated $\tau$ are compared in Table \ref{tab_computation} in  \ref{sec:typeI-typeII-complexity}.  Type I and  Type II are roughly quadratic in $(\delta,\lambda)$. Although the rectangular matrix  $\bm{\Pi}_{\delta,\lambda,\omega}$ is larger  than $\ZZk{\delta}{\lambda}{\omega}$, since $\lambda > \delta$, the economy in lazy matrix addressing combined with the number of elementary operations  make Type II implementation more practical and efficient.

\section{\chopatt: \CHOPA\label{sec:context}}
\subsection{Pattern representation and two-level context hierarchy definitions\label{subsec:contextGeneral}}
We now introduce the pattern-based approach, borrowed from common lossless image encoders (Section~\ref{sec_contribution-extrapolation}), by providing a  contextual and hierarchical framework for \chopoly extrapolation. To this aim, the role of contexts is to cover all possible scenarios (an ontology) of signal's evolution for a hybrid dynamical system. For instance, slow and steep variations must be included, the same goes for blocky and smooth signals.

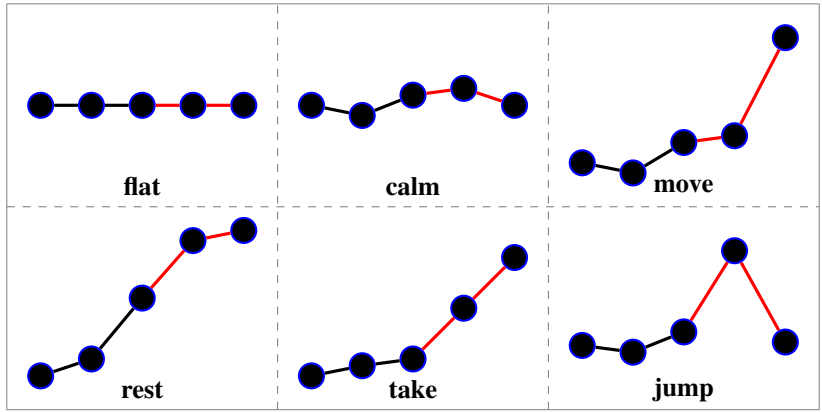
\begin{figure} [!htb]
\begin{center}
\begin{tikzpicture}[scale=0.9, place/.style={circle,draw=blue,fill=black,thick}]
  \draw[help lines] (0,0) -- (12,0);                               
  \draw[help lines] (0,0) -- (0,6);                
  \draw[help lines] (0,6) -- (12,6);  
	\draw[help lines] (12,0) -- (12,6); 
	\draw[gray,dashed] (0,3) -- (12,3);  
	\draw[gray,dashed] (4,0) -- (4,6);
	\draw[gray,dashed] (8,0) -- (8,6);  
  \node (n1) at (0.5,0.5) [place] {};
  \node (n2) at (1.25,0.75) [place] {};
  \node (n3) at (2,1.65) [place] {};
  \node (n4) at (2.75,2.5) [place] {};
  \node (n5) at (3.5,2.65) [place] {};
	\node (n6) at (2,0.3) {\textbf{rest}};

	\foreach \x [evaluate={\y=int(\x+1);}] in {1,...,2}
  \draw[very thick] (n\x) -- (n\y);
	\foreach \x [evaluate={\y=int(\x+1);}] in {3,...,4}
	\draw[red, very thick] (n\x) -- (n\y);
	\node (n1) at (4+0.5,0.5) [place] {};
  \node (n2) at (4+1.25,0.65) [place] {};
  \node (n3) at (4+2,0.75) [place] {};
  \node (n4) at (4+2.75,1.5) [place] {};
  \node (n5) at (4+3.5,2.25) [place] {};
	\node (n6) at (4+2,0.3) {\textbf{take}};
	
	\foreach \x [evaluate={\y=int(\x+1);}] in {1,...,2}
  \draw[very thick] (n\x) -- (n\y);
	\foreach \x [evaluate={\y=int(\x+1);}] in {3,...,4}
  \draw[red, very thick] (n\x) -- (n\y);
	\node (n1) at (8+0.5,0.95) [place] {};
  \node (n2) at (8+1.25,0.85) [place] {};
  \node (n3) at (8+2,1.15) [place] {};
  \node (n4) at (8+2.75,2.35) [place] {};
  \node (n5) at (8+3.5,1) [place] {};
	\node (n6) at (8+2,0.3) {\textbf{jump}};
	
	\foreach \x [evaluate={\y=int(\x+1);}] in {1,...,2}
  \draw[very thick] (n\x) -- (n\y);
	\foreach \x [evaluate={\y=int(\x+1);}] in {3,...,4}
  \draw[red, very thick] (n\x) -- (n\y);
  \node (n1) at (0.5,1.5+3) [place] {};
  \node (n2) at (1.25,1.5+3) [place] {};
  \node (n3) at (2,1.5+3) [place] {};
  \node (n4) at (2.75,1.5+3) [place] {};
  \node (n5) at (3.5,1.5+3) [place] {};
	\node (n6) at (2,0.3+3) {\textbf{flat}};
	
	\foreach \x [evaluate={\y=int(\x+1);}] in {1,...,2}
  \draw[very thick] (n\x) -- (n\y);
	\foreach \x [evaluate={\y=int(\x+1);}] in {3,...,4}
  \draw[red, very thick] (n\x) -- (n\y);
	\node (n1) at (4+0.5,1.5+3) [place] {};
  \node (n2) at (4+1.25,1.35+3) [place] {};
  \node (n3) at (4+2,1.65+3) [place] {};
  \node (n4) at (4+2.75,1.75+3) [place] {};
  \node (n5) at (4+3.5,1.5+3) [place] {};
	\node (n6) at (4+2,0.3+3) {\textbf{calm}};
	
	\foreach \x [evaluate={\y=int(\x+1);}] in {1,...,2}
  \draw[very thick] (n\x) -- (n\y);
	\foreach \x [evaluate={\y=int(\x+1);}] in {3,...,4}
  \draw[red, very thick] (n\x) -- (n\y);
	\node (n1) at (8+0.5,0.65+3) [place] {};
  \node (n2) at (8+1.25,0.5+3) [place] {};
  \node (n3) at (8+2,0.95+3) [place] {};
  \node (n4) at (8+2.75,1.05+3) [place] {};
  \node (n5) at (8+3.5,2.5+3) [place] {};
	\node (n6) at (8+2,0.3+3) {\textbf{move}};
	
	\foreach \x [evaluate={\y=int(\x+1);}] in {1,...,2}
  \draw[very thick] (n\x) -- (n\y);
	\foreach \x [evaluate={\y=int(\x+1);}] in {3,...,4}
  \draw[red, very thick] (n\x) -- (n\y);
\end{tikzpicture}
\caption{Pattern representation for the functional context in Table~\ref{tb:context}.\label{fig_six_context}} 
\end{center}
\end{figure}

On the first level, we define a functional context that differentiates a set of typical patterns, represented in Figure~\ref{fig_six_context}. We are interested in the dynamics of the most recent samples, depicted with red links, with respect to the past behavior (black links). It defines six mutually exclusive entities, illustrated by a short name: ``flat'', ``calm'', ``move'', ``rest'', ``take'' and ``jump''. 

``Flat'' addresses steady signals. ``Calm'' represents a sufficiently sampled situation, where value increments over time remain below fixed thresholds. ``Move''  defines a formerly ``calm'' signal whose novel value changes rapidly, above a pre-defined threshold. ``Rest'' handles signals previously varying above a threshold and becoming  
``calm''. The ``take'' context addresses signals with constantly high variations. The ``jump'' pattern is a ``take'' with a sign change as for bumps and dips. They are detected in practice by comparing  consecutive differences on past samples ($d_i$, \emph{cf.}~\eqref{eq:di}), indexed by $i$, to thresholds ($\gamma_i$, \emph{cf.}~\eqref{eq:gamma}). Their formal definition is provided in 
Table~\ref{tb:context} and their adaptive selection in Section~\ref{subsec:context}.

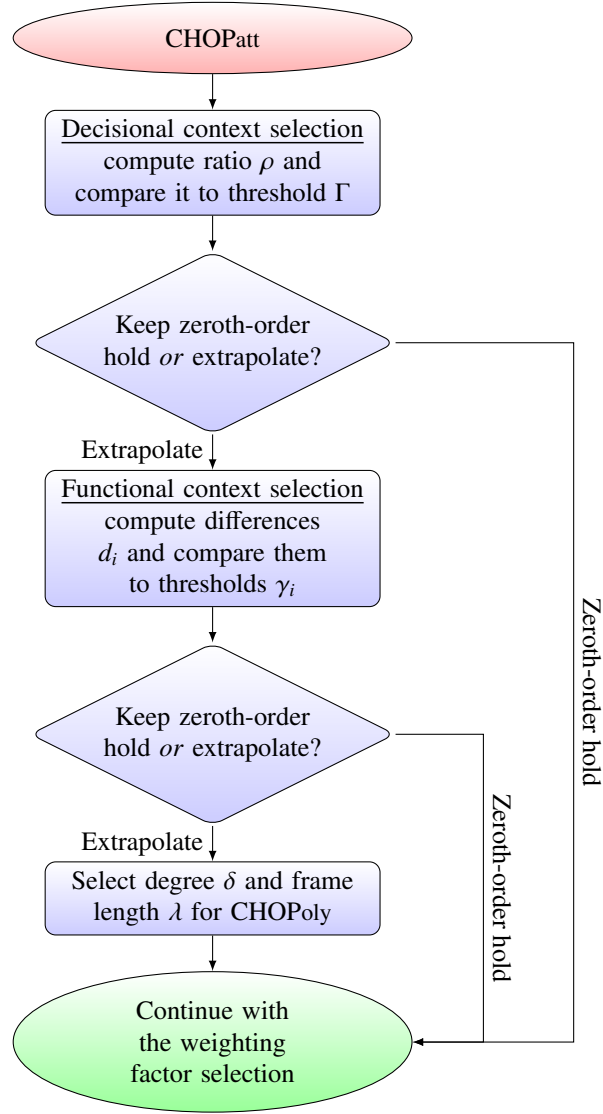
\begin{figure}[!htb]
\centering
\begin{tikzpicture}[-latex,scale=1.2]
\begin{scope}
  \matrix (chart)
    [
      matrix of nodes,
      column sep      = 8em,
      row sep         = 3ex,
      column 1/.style = {},
      column 2/.style = {}
			
    ]
    {
      |[root]|  \chopatt  \\
      |[treenode]| \underline{Decisional context selection} compute  ratio $\rho$ and compare it to  threshold $\Gamma$ \\
      |[decision]| Keep zeroth-order hold \emph{or} extrapolate?\\
      |[treenode]| \underline{Functional context selection} compute  differences $d_i$ and compare them to  thresholds $\gamma_i$ 
			\\
      |[decision]| Keep zeroth-order hold \emph{or} extrapolate? \\
      |[treenode]| Select  degree $\delta$ and  frame length $\lambda$ for \chopoly \\
      |[finish]| Continue with the weighting factor selection \\
    };
  \draw
    (chart-1-1) edge (chart-2-1)
    (chart-2-1) edge (chart-3-1)
		(chart-3-1) edge node [left]  {Extrapolate} (chart-4-1)
		(chart-4-1) edge (chart-5-1)
		(chart-5-1) edge node [left]  {Extrapolate} (chart-6-1)
		(chart-6-1) edge (chart-7-1);

   \draw (chart-3-1) -- +(4,0) |- (chart-7-1)
   node[near start,above,sloped] {Zeroth-order hold};
	\draw (chart-5-1) -- +(3,0) |- (chart-7-1)
   node[near start,above,sloped] {Zeroth-order hold};
	\end{scope}
\end{tikzpicture}
\caption{Hierarchical and functional context selection flowchart.\label{flowchart_context}} 
\end{figure}

On the second  level, we define a meta- or decisional context that determines if  extrapolation would be beneficial or detrimental. Indeed, a major difficulty for hybrid complex systems resides in sharp and fast variations in signal patterns induced by stiffness and discontinuities.
We thus detect if the signal can be characterized by a seventh ``cliff'' pattern, for which functional \chopoly prediction would fail, by taking into account the importance of the signal's amplitude. This decision context is  detected by comparing a ratio $\rho$ of differences based on past samples \eqref{eq:ratio} to a threshold (denoted by $\Gamma$, \emph{cf.}~\eqref{eq:Gamma}).
The above seven patterns  define a hierarchical context selection, composed of the decisional context embedding the functional context on a second level, summarized in Figure~\ref{flowchart_context}.

\subsection{Functional context  selection\label{subsec:context}}
As a concrete example, used in the remaining of our work, we propose two simple measures of variation based on the last three samples only: $u_{0}$, $u_{-1}$ and $u_{-2}$ (with the  zero-reindexing convention from Section \ref{smallpox_notations}), the last and previous differences:
\begin{equation}\label{eq:di}
\begin{array}{ccc}
d_0 = u_0 - u_{-1}& \operatorname{and}& d_{-1}= u_{-1} - u_{-2}\,.
\end{array}
\end{equation}

Their absolute values are compared with  thresholds $\gamma_0$ and $\gamma_{-1}$, respectively, defined in~\eqref{eq:gamma}. 
To build the different contexts, three complementary conditions are defined:
\begin{itemize}
	\item $O$ if $|d_i| = 0$;
	\item $C_i$ if $0 < |d_i| \leq \gamma_i$;
	\item $\ol{C_i}$ if $|d_i| > \gamma_i$.
\end{itemize}

Table~\ref{tb:context} formally defines  the six entities from the functional context and presents examples of ``default'' $\omega$-parametrized  \chopoly families. 
\begin{table}[!htb]\centering
\ra{1.3}
\begin{tabular}{lccccc}\toprule
n(ame) & \# & $|d_{-1}|$ & $|d_{0}|$&$d_{-1}.d_{0}$ & ($\delta,\lambda,\omega$)\\
\midrule
f(lat)& $0$ & $O$ & $O$ & $O$ &($0,1,.$)\\
c(alm) & $1$ & $C_1$ & $C_2$ &any &($2,5,.$)\\
m(ove) & $2$ & $C_1$ & $\ol{C_2}$ &any &($0,1,.$)\\
r(est) & $3$ & $\ol{C_1}$ & $C_2$ & any &($0,2,.$) \\
t(ake) & $4$ & $\ol{C_1}$ & $\ol{C_2}$ & $> 0$&($1,3,.$)\\
j(ump) & $5$ &$\ol{C_1}$ & $\ol{C_2}$ & $< 0$&($0,1,.$)\\
\bottomrule
\end{tabular}
\caption{Functional context: entities definition and associated prediction.\label{tb:context}}
\end{table}
Since the ``flat'' context addresses steady signals, a mere ZOH suffices, hence $P_{0,1,\omega}$. The ``calm'' context represents smooth signals, in this case, it could be approximated by a quadratic polynomial, for instance $P_{2,5,\omega}$. For the ``flat'' and ``jump'' contexts,  an additional procedure  consists in resetting the extrapolation to prevent inaccurate prediction. For example, when context $1$ is chosen just after context $4$, the quadratic extrapolation $P_{2,5,\omega}$ requires $5$ valid samples, whereas the last $3$ only are relevant, justifying our finite-length frames option.
\par
The choice of  thresholds $\gamma_0$ and $\gamma_{-1}$ is potentially crucial. For instance, fixed values may reveal inefficient under important amplitude or scale variation in signals. Hence, we have chosen to update them adaptively, based on the statistics of a past frame $\{u_{1-\Lambda},\ldots,u_{-3}\}$ ($\Lambda$ denotes the maximum frame size). 
\par
With excessively low thresholds, high-order extrapolations would  rarely be chosen, losing the benefits of prediction. Overly high thresholds would in contrast suffer from any unexpected jump or noise. As  contexts are based on backward derivatives, we have used in the simulations presented here the mid-range statistical estimator of their absolute values. This amounts to set:
\begin{eqnarray} \label{eq:gamma}
\displaystyle \gamma_0 = \gamma_{-1} = \frac{1}{2}\max_{i\in [{1-\Lambda},\ldots,-3]}(|u_i - u_{i+1}|)\,, 
\end{eqnarray}
which appeared to be sufficiently robust in our test-cases.
\subsection{Next communication period: weighting factor and decisional context selection\label{subsec:weighting}}
\begin{figure} [!htb]
\begin{center}
\includegraphics[width=0.7\columnwidth]{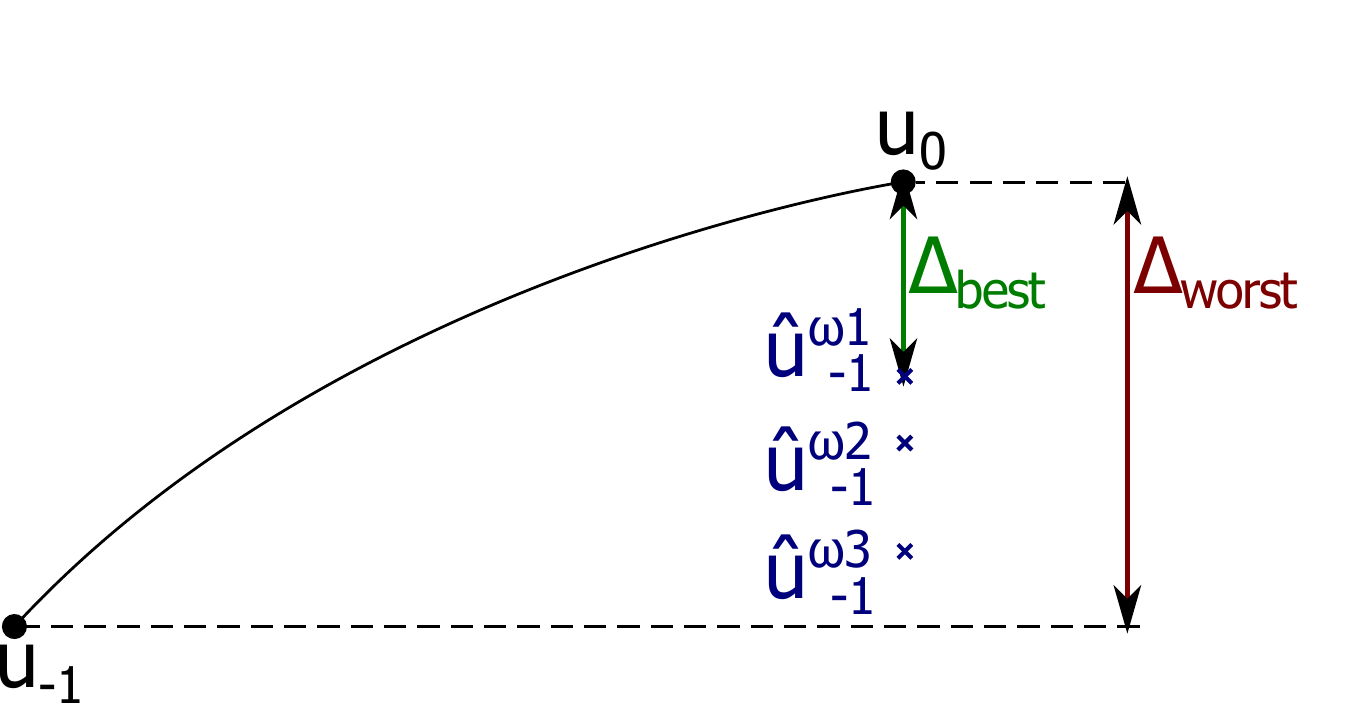} 
\caption{Illustration of two error measures:  $\Delta_{\operatorname{worst}}$ when there is no extrapolation and $\Delta_{\operatorname{best}}$ for the best prediction pattern.\label{fig_diff}} 
\end{center}
\end{figure}

To further improve our algorithm in~\cite{BenKhaled_A_2014_p-modelica_context-based_pessfmcsufmi} and to decrease even more  prediction induced integration errors, the oblivious weighting factor from~\eqref{eq_weighting} makes the algorithm more subtly aware of data freshness. 
Decisions for the next communication period are taken with respect  
to  the new updated input value $u_{0}$, compared to past predictions. It is gauged with two error measures, illustrated in Figure~\ref{fig_diff}. They assess, a posteriori, what would have been the best prediction.
First, $\Delta_{\operatorname{worst}}$ denotes the worst case scenario without extrapolation, the  predicted value being  held constant and equal to $u_{-1}$:
$$\Delta_{\operatorname{worst}}=| u_{0} - u_{-1}|.$$
Second, we estimate the best prediction pattern in retrospect, obtained by optimizing the free parameter $\omega$ in the prediction polynomial defined by the last selected context. We choose here a subset of weighting factors in $\Omega = \left \{0,\frac{1}{8},\frac{1}{4},\frac{1}{2},1,2\right \}$ (see~\ref{sec:prediction-typeII-examples} for details):
 $$\Delta_{\operatorname{best}}=\min\limits_{\omega\in \Omega}| u_{0} - \hat{u}_{-1}^{\omega}|.$$
The best weighting factor $\omega_{\operatorname{best}}$ defined in \eqref{eq:wb} is then selected during the next communication interval:
\begin{equation} \label{eq:wb}
\Delta_{\operatorname{best}}=| u_{0} - \hat{u}_{-1}^{\omega_{\operatorname{best}}}|.
\end{equation}

Regarding the selection of the decisional context called ``cliff'', we first define a ratio: 
\begin{equation} \label{eq:ratio}
\rho=\frac{\Delta_{\operatorname{best}}}{\Delta_{\operatorname{worst}}}.
\end{equation}
It specifies if there is a sharp fast variation or not by comparing it to a pre-defined threshold and then decides to extrapolate or not. 
The threshold denoted $\Gamma$ is defined as follow:
\begin{equation} \label{eq:Gamma}
0.7\leq\Gamma<1.
\end{equation}

In fact, when $\rho > \Gamma$ (e.g. with $\Gamma=\SI{90}{\percent}$), it means that the input value is only enhanced by $1-\Gamma$ (e.g. $\SI{10}{\percent}$) regarding the ``true'' value. This is the case when there is a sharp and fast variation or a weird behavior. The decisional context ``cliff'' is then selected and activated with its associated heuristic polynomial predictor $P_{\delta,\lambda,\omega}=P_{0,1,\omega}$.
On the other hand, when $\rho \leq \Gamma$, the conventional functional context table introduced in Section~\ref{subsec:context} is used.

\section{Case study\label{sec:CS}}
\subsection{Engine simulator}
In this study, a Spark Ignition (SI) RENAULT F4RT engine has been modeled  with $3$ gases (air, fuel and burned gas). It is a four-cylinder, in-line Port Fuel Injector (PFI), engine in which the engine displacement is $\SI{2000}{cm^3}$. The combustion is considered as homogeneous. The air path (AP) consists in a turbocharger with a mono-scroll turbine controlled by a waste-gate, an intake throttle and a downstream-compressor heat exchanger. This engine is equipped with two Variable Valve Timing (VVT) devices, for intake and exhaust valves, to improve the engine efficiency (performance, fuel and emissions). The maximum power is about \SI{136}{\kilo\watt} at \SI{5000}{rpm}.
\par
The F4RT engine model was developed using the ModEngine library~\cite{modengine}. ModEngine is a Modelica~\cite{Fritzson:2010} library that allows the modeling of a complete engine with diesel and gasoline combustion models.
\par
Requirements for the ModEngine library were defined and based on the already existing IFP-Engine library. The development of the IFP-Engine library was performed several years ago at ``IFP Energies nouvelles'' and it is currently used in the AMESim\footnote{www.lmsintl.com/imagine-amesim-1-d-multi-domain-system-simulation} tool. 
ModEngine contains more than $250$ sub-models. It has been developed to allow the simulation of a complete virtual engine using a characteristic time-scale based on the crankshaft angle. A variety of elements are available to build representative models for engine components, such as turbocharger, wastegate, gasoline or diesel injectors, valve, air path, Exhaust Gas Recirculation (EGR) loop etc. ModEngine is currently functional in Dymola\footnote{www.3ds.com/products/catia/portfolio/dymola}.
\par
The engine model and the split parts were imported into xMOD model integration and virtual experimentation tool~\cite{x}, using the FMI export features of Dymola. This cyber-physical system has $118$ state variables and $312$ event indicators (of discontinuities).
\subsection{Decomposition approach}
The partitioning of the engine model is performed by separating the four-cylinder from the air path, then by isolating the cylinders ($C_i$, for $i\in\{1,\ldots,4\}$) from each other. 
This kind of splitting allows for the reduction of the number of events acting on each sub-system. In fact, the combustion phase raises most of the events, which are located in the firing cylinder. The solver can process them locally during the combustion cycle of the isolated cylinder, and then enlarge its integration time-step until the next cycle.
\par
From a thermodynamic point of view, the cylinders are weakly coupled, but a mutual data exchange does still exist between them and the air path.
\par
The dynamics of the air path is slow (it produces slowly varying outputs to the cylinders, e.g. temperature) compared to those of the cylinders (they produce fast outputs to the air path, e.g. torque). Besides, unlike cylinders outputs, most air path outputs are not a direct function of the air path inputs (they are called Non Direct Feedthrough (NDF) outputs). This results in choosing the execution order of the split model from the air path to the cylinders (in accordance with the analysis of the behavior of NDF to Direct Feedthrough (DF) in~\cite[Chapter 9]{Benkhaled_2014}).
\par
The model is split into $5$ components and governed by a basic controller denoted CTRL. It gathers $91$ inputs and $98$ outputs. 

\section{Tests and results\label{sec:results}}

Tests are performed on a platform with $\SI{16}{GB}$ RAM and an ``Intel Core i7'' 64-bit processor, running $4$ cores ($8$ threads) at $\SI{2.70}{\giga\hertz}$. 
\subsection{Reference simulation}
The model validation is based on the observation of some quantities of interest as the pressure, the gas mass fraction, the enthalpy flow rate, the torque, etc. These outputs are computed using LSODAR \footnote{Short for Livermore Solver for Ordinary Differential equations with Automatic method switching for stiff and nonstiff problems, and with Root-finding \cite{Hindmarsh}.}, a variable time-step solver with a root-finding capability that detects the events occurring during the simulation. It also has the ability to adapt the integration method depending on the observed system stiffness.

The simulation reference $Y_{\operatorname{ref}}$ is built from the integration of the entire engine model, the solver tolerance ($\operatorname{tol}$) being decreased until reaching stable results, which is reached for $\operatorname{tol}=10^{-7}$ (at the cost of an unacceptable slow simulation speed). 

Then, to explore the trade-offs between the simulation speed and precision, simulations are run with increasing values of the solver tolerance until reaching a desired relative integration error $\operatorname{Er}$, defined by~\eqref{eq:er}
\begin{equation}\label{eq:er}
\operatorname{Er}(\%)=\frac{100}{N}.\sum \limits_{i=0}^{N-1}\left|\frac{Y_{\operatorname{ref}}(i) - Y(i)}{Y_{\operatorname{ref}}(i)}\right|,
\end{equation}
with $N$ the number of saved points during $\SI{1}{\second}$ of simulation.
Iterative runs showed that the relative error converge to a desired error ($\operatorname{Er} \leq \SI{1}{\percent}$) for $\operatorname{tol}=10^{-4}$. The single-thread simulation of the whole engine with LSODAR and $\operatorname{tol}=10^{-4}$ provides the simulation execution time reference, to which the parallel version is compared. When using the split model, each of its $5$ components is assigned to a dedicated core and integrated by LSODAR with $\operatorname{tol}=10^{-4}$.

\subsection{Automatic detection of fast and sharp variations\label{sec_sharp-variations}}
\begin{figure}[!htb]
\centering
\begin{subfigure}{.5\textwidth}
  \centering
  \includegraphics[width=.9\linewidth]{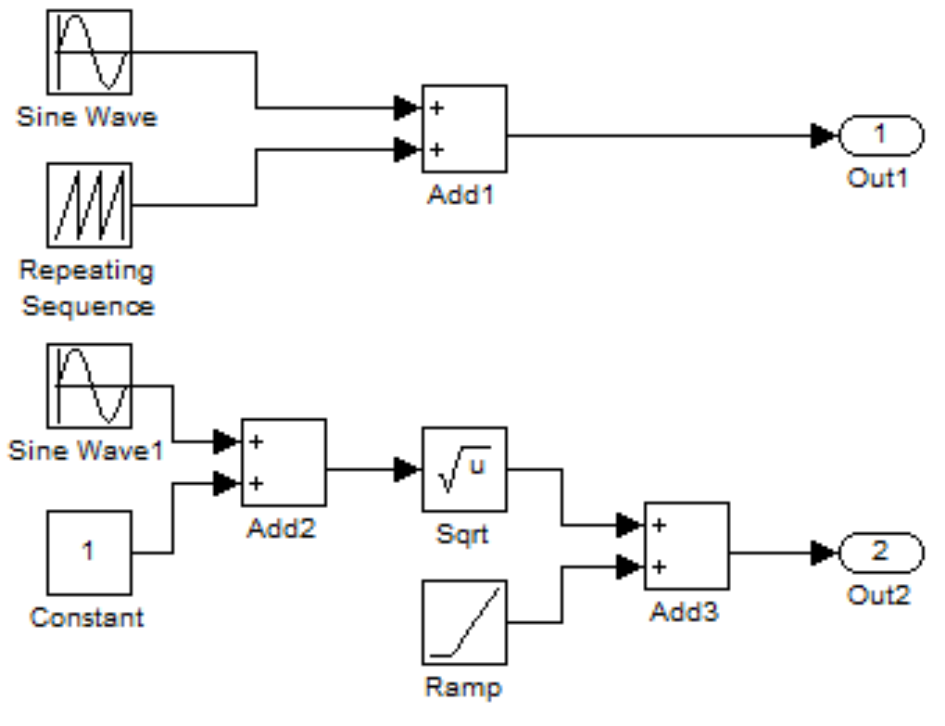}
  \caption{Construction of Out1 and Out2 in Matlab/Simulink.\label{fig:Simulink}}
\end{subfigure}
\begin{subfigure}{.5\textwidth}
  \centering
  \includegraphics[width=.9\linewidth]{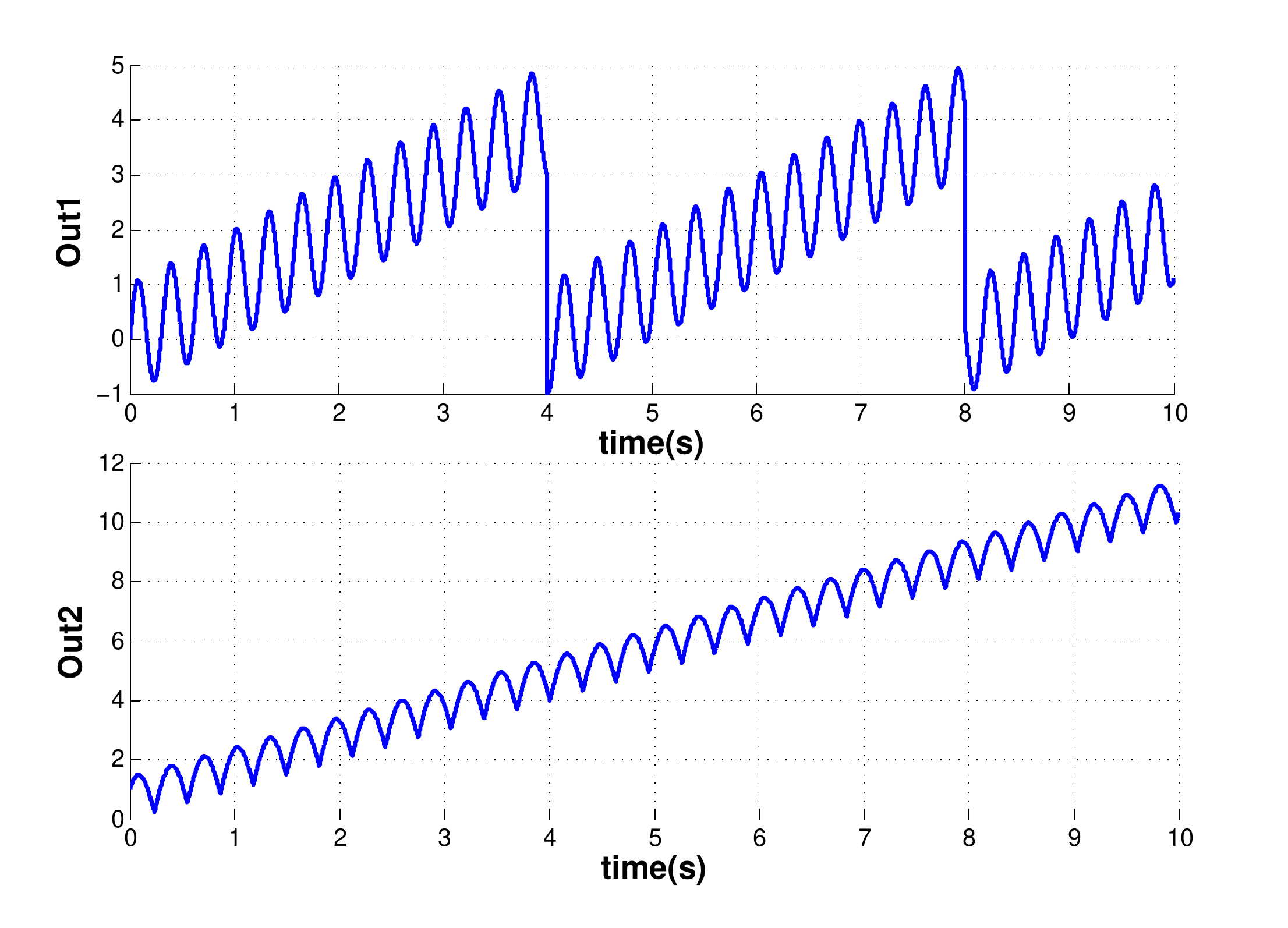}
  \caption{Illustration of Out1 and Out2.\label{fig:SigOuts}}
\end{subfigure}
\caption{A test sample.\label{fig:testSample}}
\end{figure}

\begin{figure} [!htb]
\begin{center}
\includegraphics[width=0.75\columnwidth]{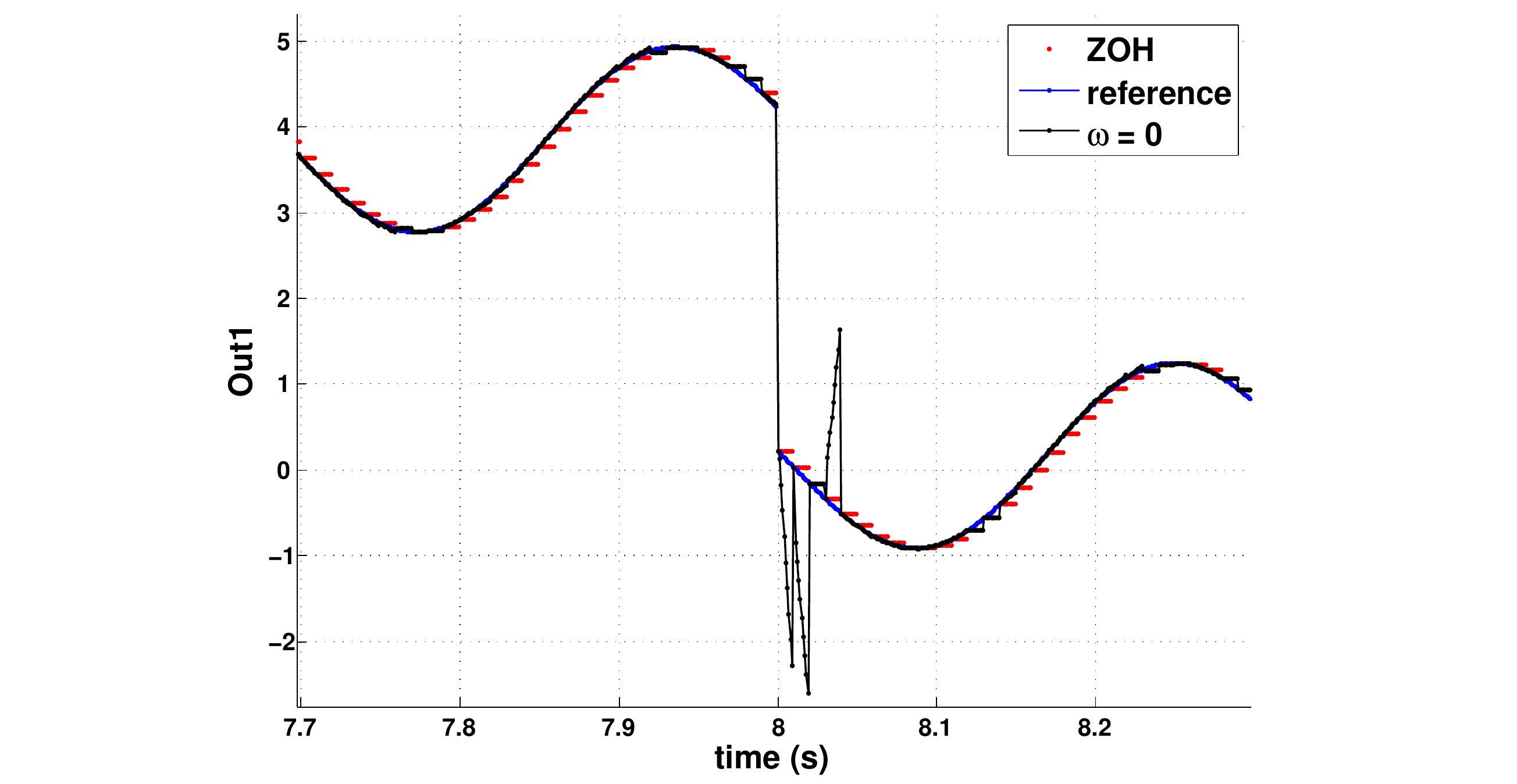}
\caption{Failure of the old extrapolation.\label{fig:jump}} 
\end{center}
\end{figure}

Adding a hierarchy of decisional  and functional contexts overcomes previous limitation in~\cite{BenKhaled_A_2014_p-modelica_context-based_pessfmcsufmi}. We illustrate with two signals denoted ``Out1'' and ``Out2''. They are built in Matlab/Simulink as shown in Figure~\ref{fig:Simulink}. 
They exhibit different variations over time, as illustrated in Figure~\ref{fig:SigOuts}. 
From Figure~\ref{fig:jump}, extrapolation of ``Out1'' fails around $t=\SI{8}{\second}$ at the sharp variation. Here extrapolation is detrimental since it increases errors instead of minimizing them.

\begin{figure} [!htb]
\begin{center}
\includegraphics[width=0.75\columnwidth]{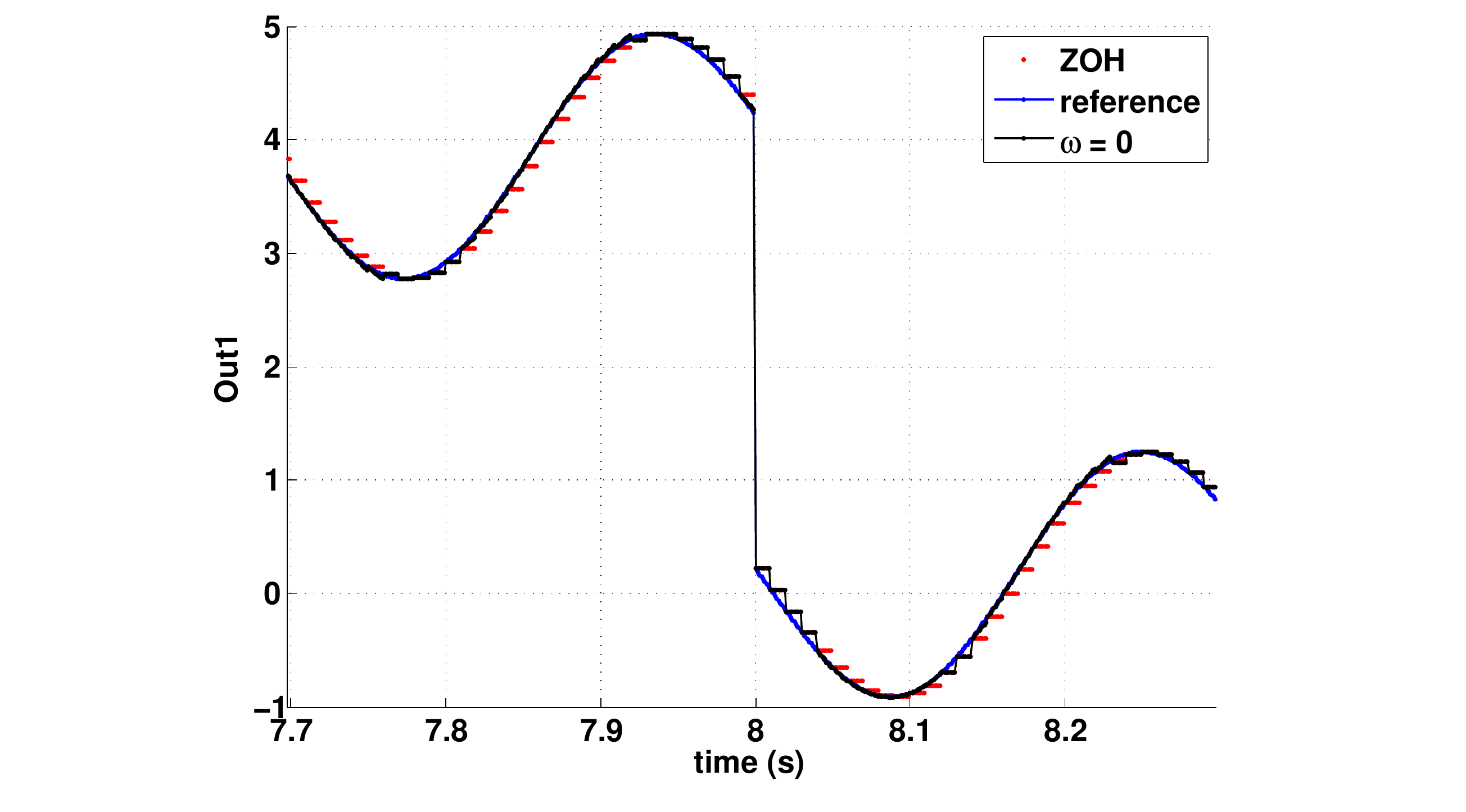}
\caption{Success of the new extrapolation.\label{fig:jump_correct}} 
\end{center}
\end{figure}

To fix this limitation on high jumps, we apply hierarchical contexts' selection to detect the ``cliff'' context. Using the ratio $\rho$ \eqref{eq:ratio} and comparing it to a threshold $\Gamma$ \eqref{eq:Gamma} equal to \SI{90}{\percent}, the improvement of extrapolation at this step is found lesser than \SI{10}{\percent}. The ``cliff'' context is then activated to avoid extrapolation. As a result, the decisional context prevents additional errors of prediction and the result of the simulation is improved as it is shown in Figure~\ref{fig:jump_correct}.

\subsection{Automatic selection of the weighting factor}
To determine the best weighting factor regarding error minimization, we first test them separately on ``Out1'' and ``Out2''. This first test is quite simple since there is no interaction between blocks, which means that there is no effect of the action/reaction of extrapolated signals on each other.
The purpose here is to show the difference between all weighting factors.

\begin{figure} [!htb]
\begin{center}
\includegraphics[width=0.75\columnwidth]{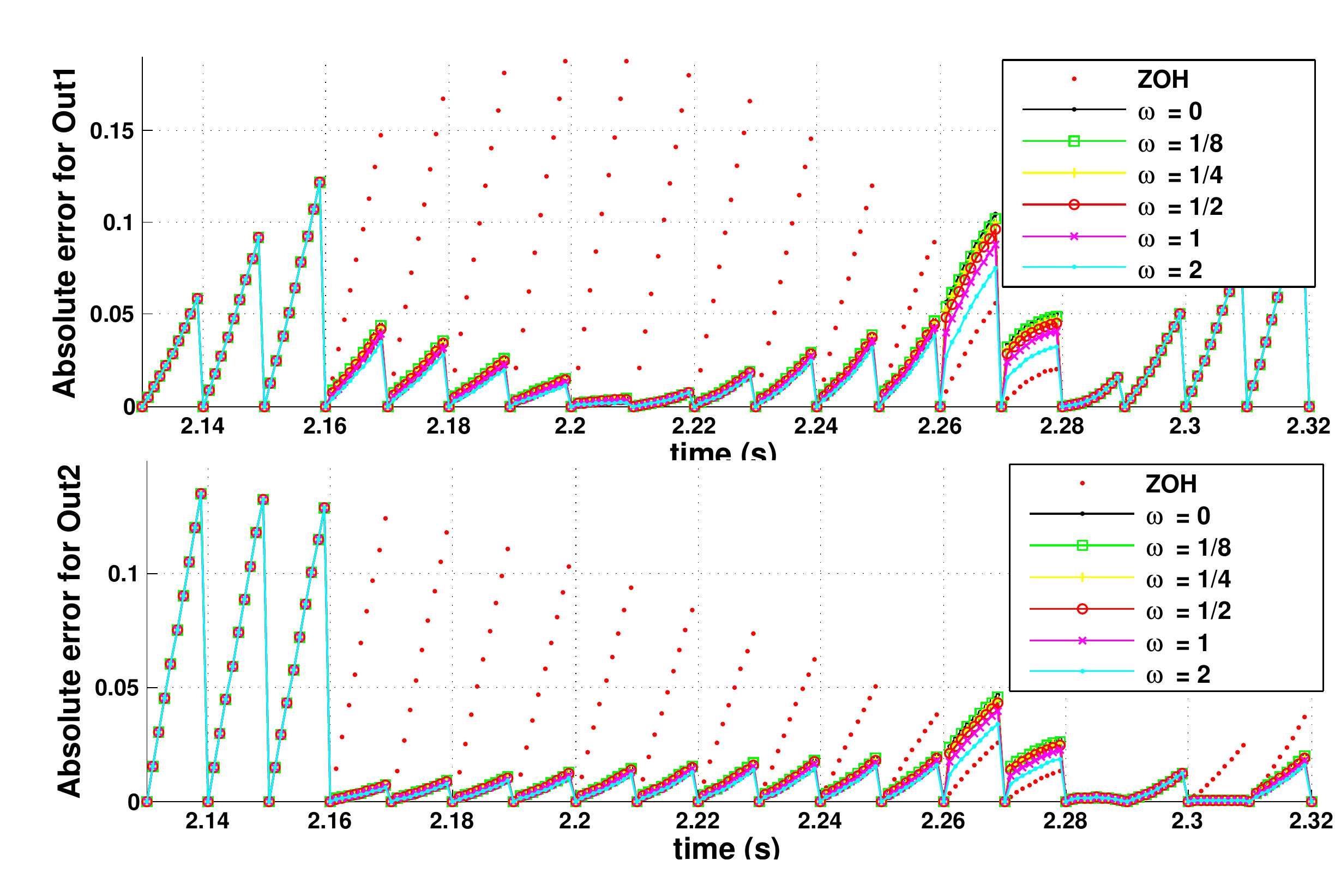} 
\caption{Behavior of the absolute error at each communication step.\label{fig:absErr}} 
\end{center}
\end{figure}

Figure~\ref{fig:absErr} shows the absolute error, which is the absolute value of the difference between the reference in Figure~\ref{fig:SigOuts} built with a small communication step $H=\SI{10}{\micro\second}$ and signals simulated with a larger communication step $H=\SI{100}{\micro\second}$, extrapolated or not. It can be inferred that the higher the weighting factor, the smaller the error.

Table~\ref{tb:err} shows the cumulative absolute error on a long simulation run (during \SI{10}{\second}). It confirms that the weighting factor $\omega=2$ is the best regarding error reduction. In fact, the extrapolation is enhanced from $\omega=0$ to $\omega=2$ by reducing the error of prediction by \SI{20.10}{\percent} for ``Out1'' and by \SI{11.39}{\percent} for ``Out2''.
\begin{table}[!htb]\centering
\ra{1.3}
\begin{tabular}{cccccc}\toprule
\multirow{2}{*}{Type}&\multicolumn{2}{c}{Error}&&\multicolumn{2}{c}{Improvement} \\
\cmidrule{2-3} \cmidrule{5-6}
&Out1&Out2&&Out1&Out2\\
\midrule
ZOH&0.572&0.399&&---&---\\
$\omega=0\phantom{/8}$&0.204&0.158&&\SI{64.34}{\percent}&\SI{60.40}{\percent}\\
$\omega=1/8$&0.201&0.157&&\SI{64.86}{\percent}&\SI{60.65}{\percent}\\
$\omega=1/4$&0.198&0.155&&\SI{65.38}{\percent}&\SI{61.15}{\percent}\\
$\omega=1/2$&0.193&0.153&&\SI{66.27}{\percent}&\SI{61.65}{\percent}\\
$\omega=1\phantom{/8}$&0.182&0.148&&\SI{68.18}{\percent}&\SI{62.91}{\percent}\\
$\omega=2\phantom{/8}$&0.163&0.140&&\SI{71.50}{\percent}&\SI{64.91}{\percent}\\
\bottomrule
\end{tabular}
\caption{Cumulative absolute error during \SI{10}{\second} and relative improvement.\label{tb:err}}
\end{table}

\begin{figure} [!htb]
\begin{center}
\includegraphics[width=0.7\columnwidth]{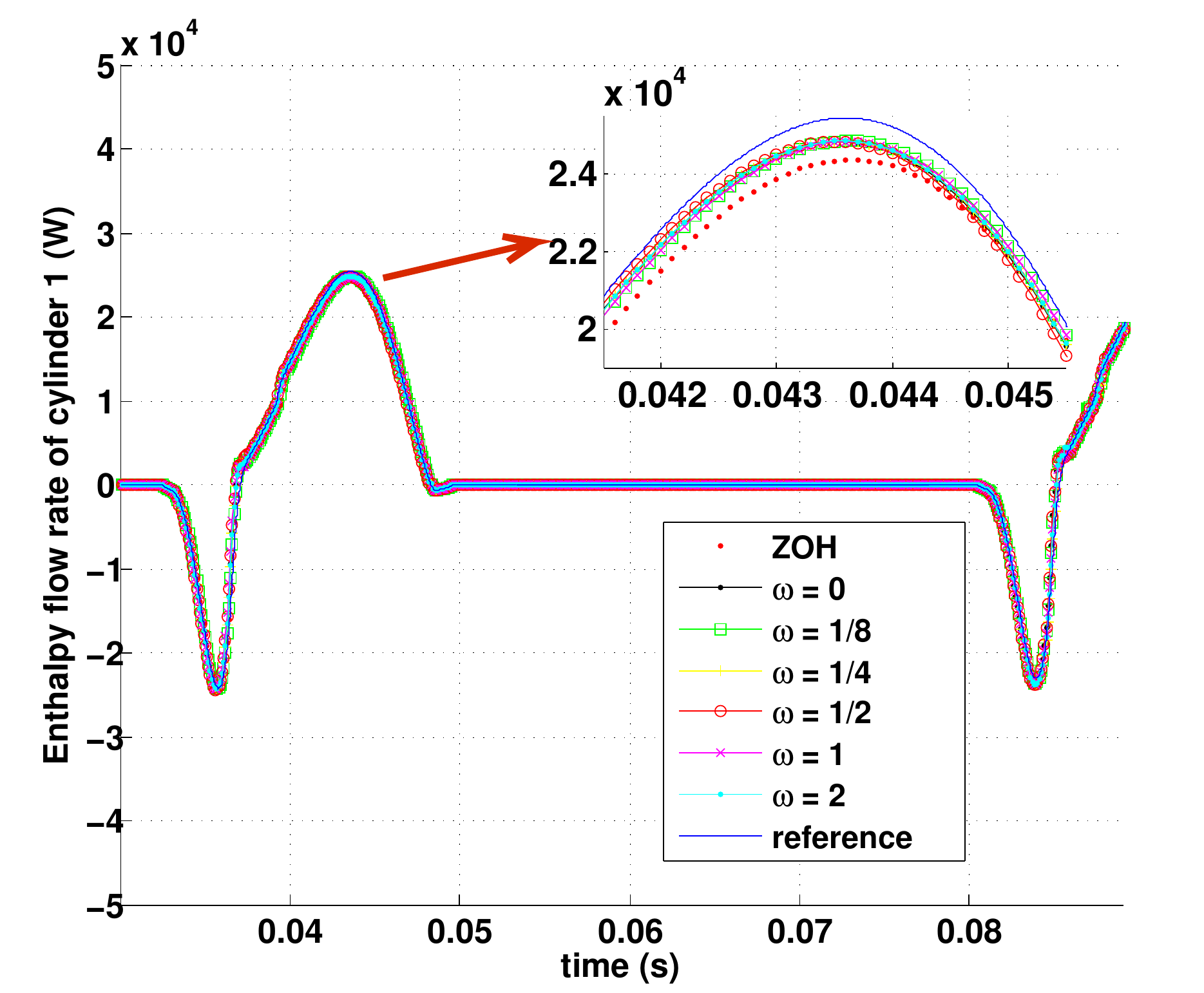}
\caption{Enthalpy flow rate output of cylinder~1.\label{fig:outCYL}} 
\end{center}
\end{figure}

The same experience is now applied on the F4RT engine model and extrapolation with different weighting factors is applied separately on all engine inputs. Figure~\ref{fig:outCYL} represents one of the cylinder~1 outputs, ``the enthalpy flow rate'', for the different extrapolation modes.

\begin{figure} [!htb]
\begin{center}
\includegraphics[width=0.75\columnwidth]{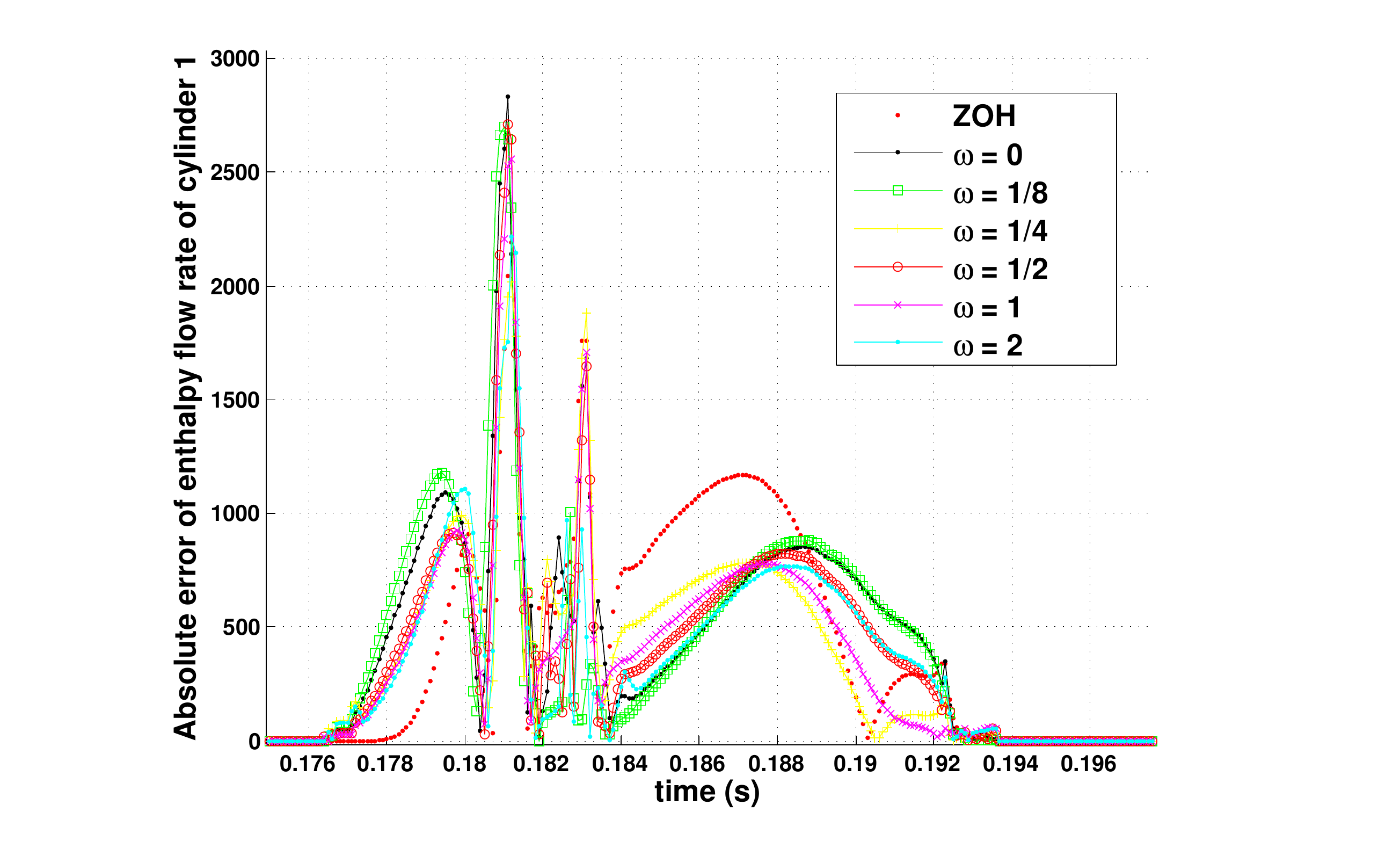}
\caption{Absolute error of enthalpy flow rate output of cylinder~1.\label{fig:ERRoutCYL}} 
\end{center}
\end{figure}

We notice more clearly in Figure~\ref{fig:ERRoutCYL} that for each communication step, there is a different best weighting factor that minimizes the absolute error. Besides, the computation of the cumulative integration error, during a long simulation run, shows that there is no unique best weighting factor. 

The weighting factor $\omega$ is then chosen dynamically during the simulation as described in Section~\ref{subsec:weighting}. At each communication step, the weighting factor that minimizes the last error is selected and used for the current step. 

\begin{figure} [!htb]
\centering
\begin{subfigure}{.5\textwidth}
  \centering
  \includegraphics[width=1.03\linewidth]{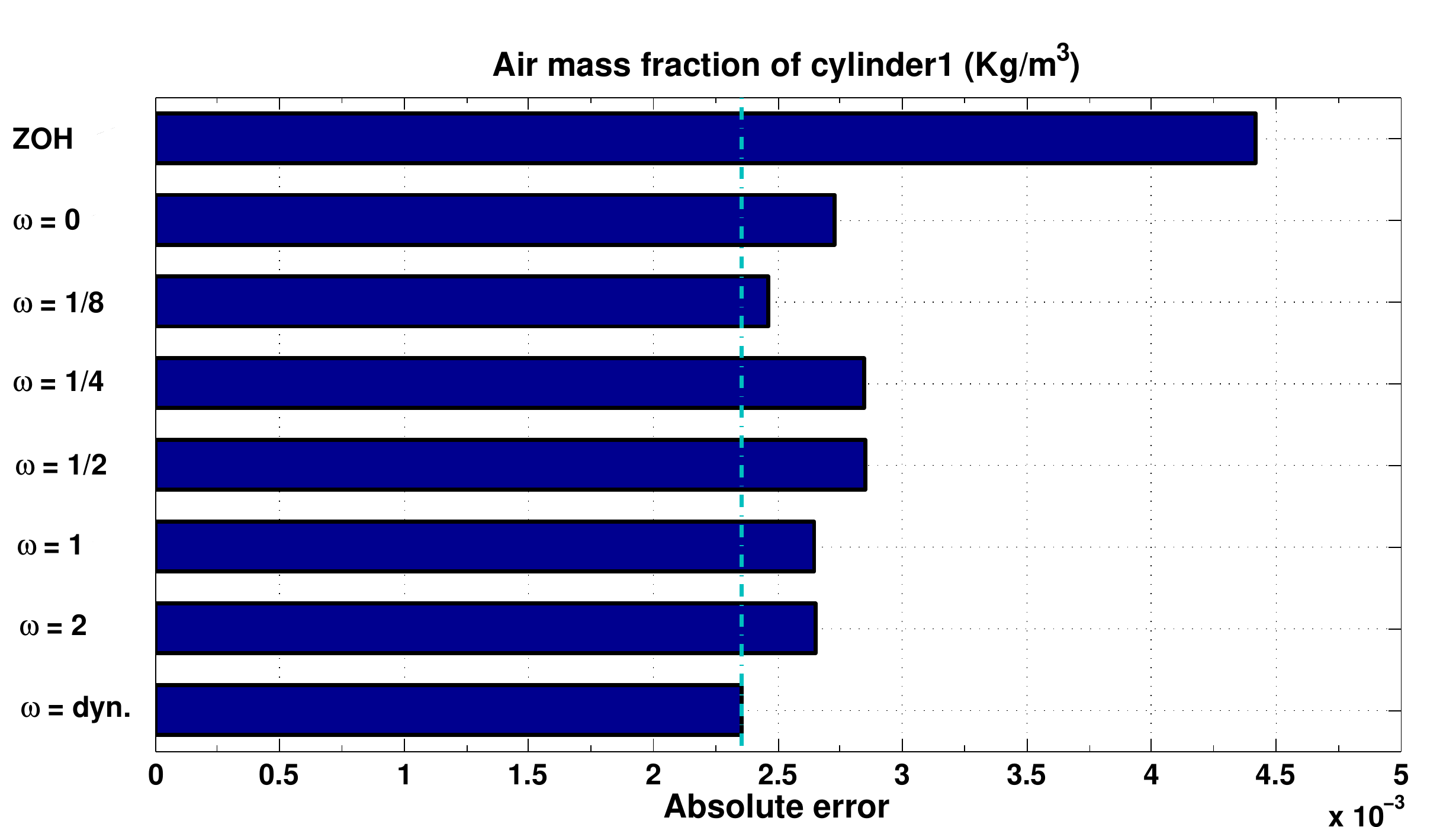}
  \caption{Air mass fraction.\label{fig:err1}}
\end{subfigure}%
\begin{subfigure}{.5\textwidth}
  \centering
  \includegraphics[width=1.03\linewidth]{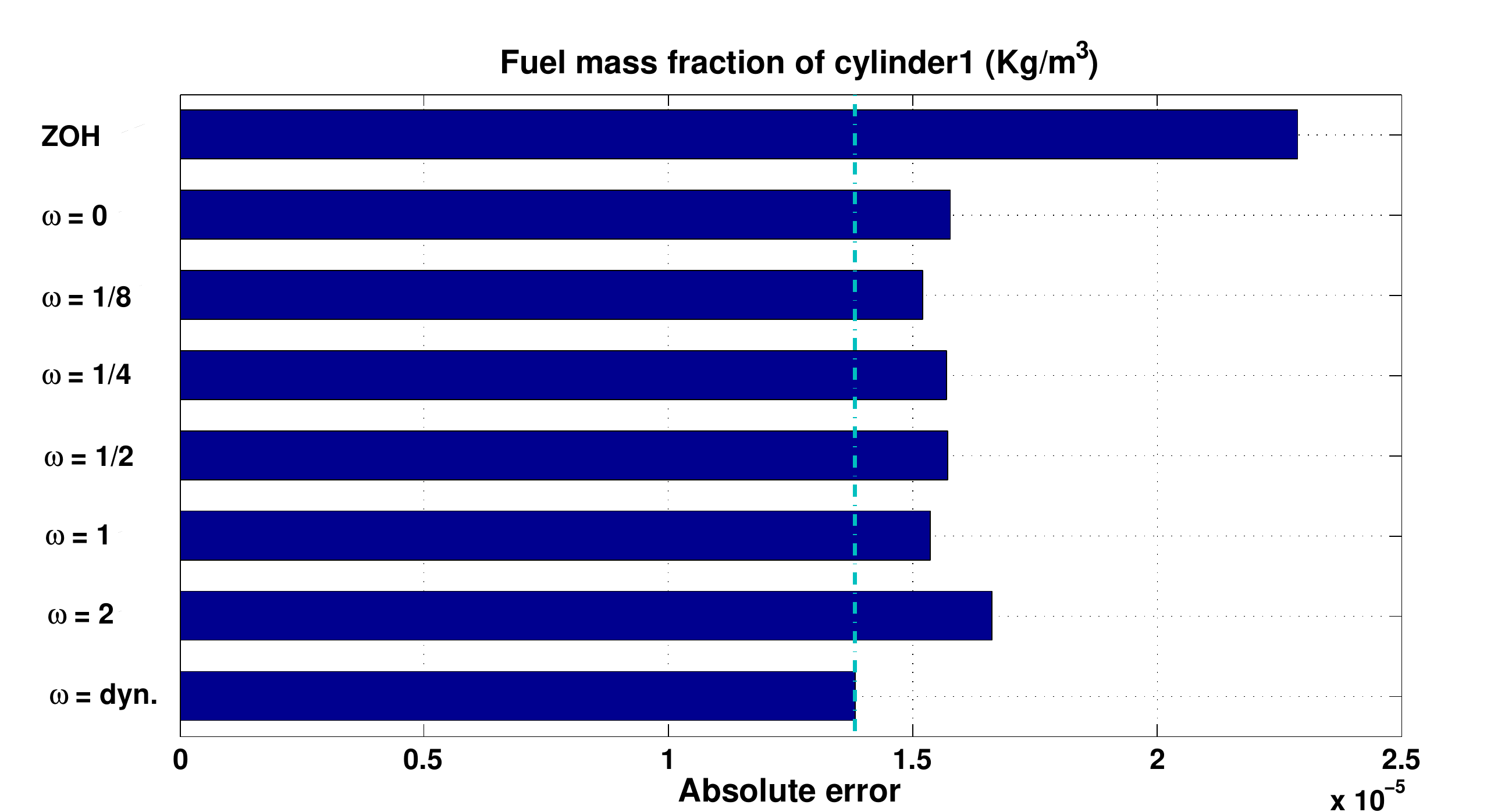}
  \caption{Fuel mass fraction.\label{fig:err2}}
\end{subfigure}
\caption{Absolute error in cylinder~1.\label{fig:test}}
\end{figure}

Thanks to dynamic error evaluation and weighting factor selection, the cumulative integration error is almost always the lowest one for the different outputs as  shown in Figures~\ref{fig:err1} and~\ref{fig:err2}. However, the worst enhancement of the error depends on the output, for instance it is obtained with $\omega=\frac{1}{2}$ for the air mass fraction of the cylinder~1 and with $\omega=2$ for the fuel mass fraction of the cylinder~1. This confirms that for complex coupled systems, there is no unique best weighting factor, hence the necessity and the usefulness of combining different ones. 

\begin{figure} [!htb]
\begin{center}
\includegraphics[width=0.7\columnwidth]{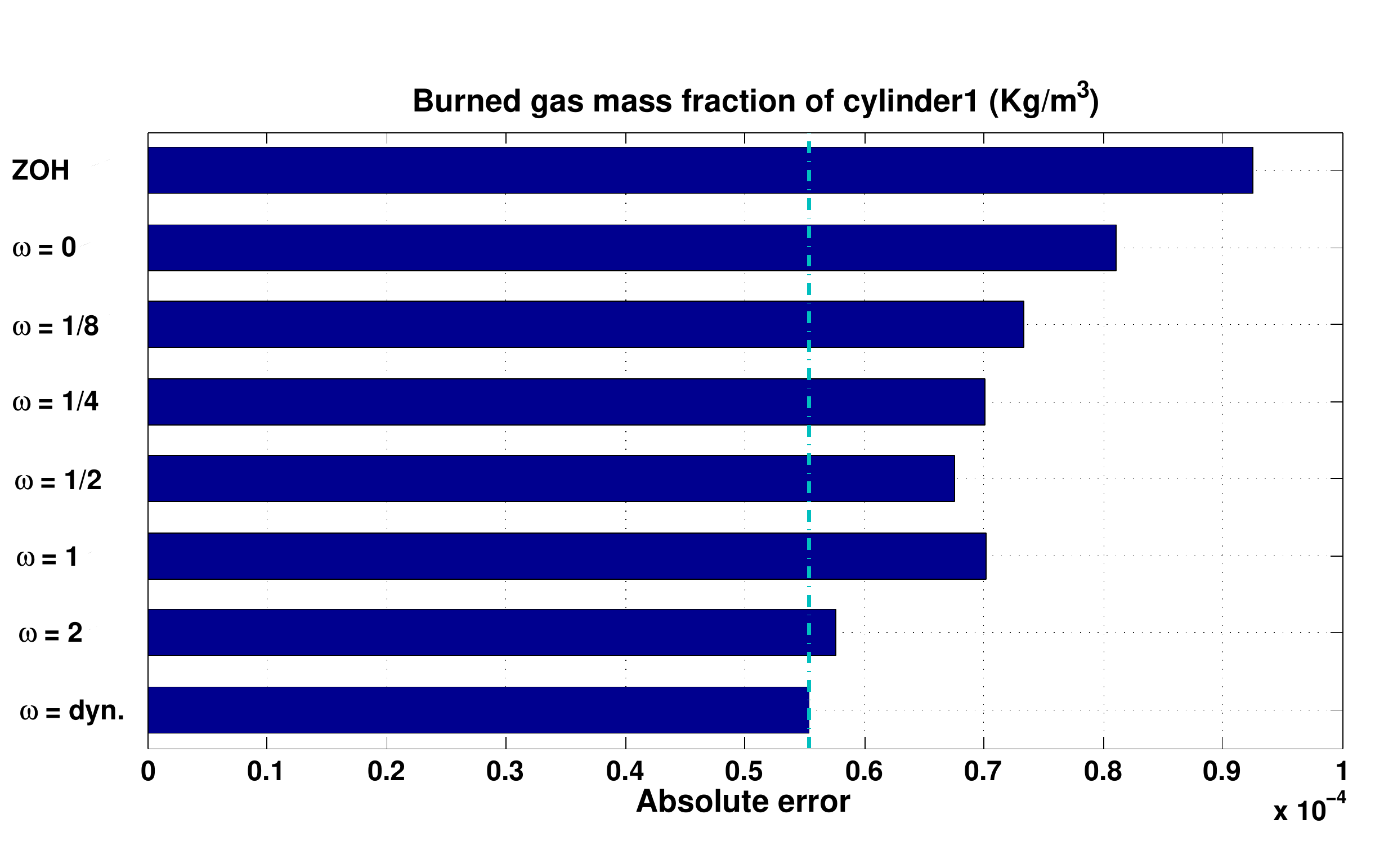} 
\caption{Absolute error of the burned gas mass fraction of cylinder~1.\label{fig:err3}} 
\end{center}
\end{figure}

Besides, for the ``burned gas mass fraction'' of cylinder~1 (see Figure~\ref{fig:err3}), the dynamic weighting factor selection decreases the error of prediction by \SI{32}{\percent} compared to the previous work (with $\omega=0$, in~\cite{BenKhaled_A_2014_p-modelica_context-based_pessfmcsufmi}) as well as the simulation error by \SI{40}{\percent} compared to the non-extrapolated signal. 

Regarding now the achievement on the simulation speed-up, Table~\ref{tb:sp} shows the acceleration compared with the single-threaded reference. Firstly, the speed-up is supra-linear w.r.t. the number of cores when the model is split into $5$ threads integrated in parallel on $5$ cores. Indeed, the containment of events detection and handling inside small sub-systems allows for solvers accelerations, enough to over-compensate the multi-threading costs. 
Secondly, model splitting combined with enlarged communication steps, from $H=\SI{100}{\micro\second}$ to $H=\SI{250}{\micro\second}$, allows around +\SI{12.5}{\percent} extra speed-up. Unfortunately this extra speed-up is obtained at the cost of relative error increase (e.g. +\SI{341}{\percent} for the fuel density). Thirdly, the combination of model splitting with expanded communication steps (use of $H=\SI{250}{\micro\second}$) as well as \choptrey allows to keep the same extra speed-up while decreasing the relative error to values close to, or below, those measured with $H=\SI{100}{\micro\second}$ and ZOH. 
We can conclude that the enhancement brought with \choptrey allows for improved performance on  both sides: simulation time and results' accuracy. 

\begin{table}[!htb]
\begin{center}
\ra{1.3}
\begin{tabular}{ccccc}
\toprule
 \multirow{2}{*}{Communication }&\multirow{2}{*}{Prediction}&\multirow{2}{*}{Speed-up }&\multicolumn{2}{c}{Relative error variation}\\
\cmidrule{4-5}
time&&factor&Burned gas density&Fuel density\\
\midrule
$\SI{100}{\micro\second}$& ZOH&8.9&--& --\\
\rule[1pt]{0pt}{18pt} 
 \multirow{2}{*}{$\SI{250}{\micro\second}$}& ZOH& 10.01& +\SI{7}{\percent}&+\SI{341}{\percent} \\
&\choptrey&10.07&\SI{-26}{\percent}&+\SI{21}{\percent}\\
\bottomrule
\end{tabular}
\caption{\choptrey performance: speed-up vs. accuracy. The speed-up factor is compared with single-threaded reference. The relative error variation is compared with ZOH at $\SI{100}{\micro\second}$.\label{tb:sp}}
\end{center}
\end{table}

\section{Summary and perspectives\label{conc}}

The main objective of \choptrey is to provide a framework for hybrid dynamical systems co-simulation speed-up~\cite{Broman_D_2014_tr_requirements_hc} based on extrapolation. 
Cheaper slackened synchronization intervals are allowed by a combination of prediction and multi-level context selection.
It aims at reaching real-time simulation while preserving result accuracy.

It is implemented in combination with  model splitting and parallel simulation on a hybrid dynamical engine model. It results in effective  simulation speed-up with negligible computational overheads. In addition,  sustained or even improved simulation precision is obtained without noticeable instability. 

This work can be extended in different directions. Keeping with data extrapolation, simulated signals can be cleaned from long range trends~\cite{Ning_X_2014_j-chemometr-intell-lab-syst_chromatogram_bedusbeads}, to better detect subtle behavioral modifications, and subsequently adapt detection thresholds. They can be processed on different time grids~\cite{Gonzalez_F_2010_j-multibody-syst-dyn_effect_mcsteamsd} with multi-scale techniques~\cite{Chaux_C_2007_j-ieee-tit_noise_cpdtwd}. Acting as local pseudo-derivatives, the latter can decompose signals into morphological components such as polynomials trends, singularities and oscillations.   This would allow improvements in context assignment by measuring sharp variations and spurious events with data-relative sparsity metrics~\cite{Repetti_A_2015_j-ieee-spl_euclid_tsbdsl1l2r}.

Moreover, the discrimination of cliff behaviors could be further improved by using knowledge of the plant model, allowing to discard out-of-bound values, e.g., non-negative variables. 

Finally, simulation results suggest that widening the communication steps is an important source for integration acceleration. Beyond equidistant communication grids, adaptive, context and/or error-based closed-loop control of communication steps~\cite{Schierz_T_2012_p-modelica_co-simulation_csscfmicma,Sadjina_S_2016_PREPRINT_energy_cpbcsniasscee} is a promising research objective.

\section{Acknowledgments}
This work was supported by the ITEA2 project MODRIO\footnote{Model Driven Physical Systems Operation}, and funded in part by the ``Direction G\'en\'erale des Entreprises'' of the French Ministry of Industry.

\appendix
\section{Complements on extrapolation\label{sec:appendix}}

\subsection{Toy parabolic \chopoly extrapolation formulae 
\label{sec:prediction-parabola}}
We estimate the best fitting parabola (i.e. $\delta= 2$) with a uniform weighting ($\omega=0$): $u(t) = a_{\delta} + a_{\delta-1}t+a_{\delta-2}t^2$ to approximate the
set of discrete samples
$\{u_{1-\lambda},u_{2-\lambda},\ldots,u_{0}\}$. 
We consider here ``uniform'' weighting $w_l = 1$ for $1-\lambda \leq l \leq 0$ (i.e. with weighting factor $\omega= 0$).
The prediction polynomial $P_{2,\lambda,0}$ is defined by the vector of polynomial coefficients: 
$\bm{a}_2 = [a_2,a_1,a_0]^T$.
These coefficients are determined, in the least-squares sense~\cite{Stigler_S_1981_j-ann-stat_gauss_ils}, by minimizing the squared  or quadratic prediction error \eqref{eq:error-d-l-w}:
$$ e(\bm{a}_2) = 
\sum_{l=1-\lambda}^{0} 1 \times \left(  u_{l} - ( a_2+ a_1 l+a_0 l^2) \right)^2 \,.$$  
Here  indices $l$ are non-positive, i.e. between $1-\lambda$ and $0$. The minimum error is obtained by solving the following system of equations (zeroing the derivatives with respect to each of  the free variables $a_i$):
\[
\forall i \in \{0,1,2\},\quad \displaystyle \frac{\partial e(\bm{a}_2) }{\partial a_i}= 0 \,,\\ 
\]
namely:
\begin{equation}
\left\{
\begin{array}{l}
\displaystyle 
\sum^0_{l=1-\lambda} l^0 \left(  u_{l} - ( a_2l^0+a_1 l^1+a_0l^2) \right)  = 0 \,,\\
\displaystyle 
\sum^0_{l=1-\lambda} l^1 \left(  u_{l} - ( a_2l^0+a_1 l^1+a_0l^2) \right)  = 0 \,,\\
\displaystyle \,
\sum^0_{l=1-\lambda} l^2 \left(  u_{l} - ( a_2l^0+a_1 l^1+a_0l^2) \right)  = 0\,.
\end{array} 
\right.
\label{eq:xdef}
\end{equation}
System  \eqref{eq:xdef} may be rewritten as:
\[
\left\{
\begin{array}{l}
\displaystyle \sum^0_{1-\lambda}  u_{l} = a_2 \sum^0_{1-\lambda} l^0   + a_1 \sum^0_{1-\lambda} l^1   + a_0 \sum^0_{1-\lambda} l^2\,,   \\
\displaystyle \sum^0_{1-\lambda} l  u_{l} = a_2 \sum^0_{1-\lambda} l^1   + a_1 \sum^0_{1-\lambda} l^2   + a_0 \sum^0_{1-\lambda} l^3 \,,  \\
\displaystyle \sum^0_{1-\lambda} l^2  u_{l} = a_2 \sum^0_{1-\lambda} l^2   + a_1 \sum^0_{1-\lambda} l^3   + a_0 \sum^0_{1-\lambda} l^4\,.   \\
\end{array} 
\right.
\]
Closed-form expressions exist for the sum of powers $\Sk{d}{\lambda}$, involving Bernoulli sequences~\cite{DeBruyn_G_1994_j-fibonacci-q_for_f12p3pnp}. For instance, up to the $4^{\operatorname{th}}$ power:
\begin{itemize}
\item $\Sk{0}{\lambda} = \lambda$; 
\item $\Sk{1}{\lambda} = (\lambda-1)\lambda/2$;
\item  $\Sk{2}{\lambda} = (\lambda-1)\lambda(2\lambda-1)/6$;
\item  $\Sk{3}{\lambda} = (\lambda-1)^2\lambda^2/4$; 
\item  $\Sk{4}{\lambda} = (\lambda-1)\lambda(2\lambda-1)(3\lambda^2-3\lambda-1)/30$. 
\end{itemize}
Let $\mk{d}{\lambda} = \mmk{d}{\lambda}{0} =\sum_{l=0}^{\lambda-1} {l }^{d}  u_{-l}$ (here  indices $l$ are positive) denote the $d^{\operatorname{th}}$ moment\footnote{Definition: the $d^{\operatorname{th}}$ moment of a real function $f(t)$ about a constant  $c$ is usually defined as: $M_d = \int_{-\infty}^{\infty} (t-c)^d f(t)dt$. The $m_{d}$s may be interpreted as  discrete versions of one-sided  moments about $t=0$ of the discrete function $u(t)$; alternatively --- \emph{cf.} definitions for $\Sk{d}{\lambda}$ --- the moments are sort of weighted (by $u_l$) sum-of-powers.} 
of the samples $u_{i}$,
and $\bm{m}_{2,\lambda}$ the vector of moments $ [\mk{0}{\lambda},-\mk{1}{\lambda},\mk{2}{\lambda}]^T$.
We now form   the Hankel matrix 
\Zk{2}{\lambda}
of sums of powers (depending on $\delta=2$ and $\lambda$):
$$ \Zk{2}{\lambda} =
\left[ \begin{array}{rrr}
  \Sk{0}{\lambda} & -\Sk{1}{\lambda} &  \Sk{2}{\lambda} \\
  -\Sk{1}{\lambda} & \Sk{2}{\lambda} & -\Sk{3}{\lambda} \\
  \Sk{2}{\lambda}    & -\Sk{3}{\lambda}    & \Sk{4}{\lambda} 
  \end{array}\right] \,.
$$
The system in~\eqref{eq:xdef} rewrites:
$$ \left[ \begin{array}{r} \mk{0}{\lambda} \\ -\mk{1}{\lambda} \\ \mk{2}{\lambda} \end{array}\right] = 
\left[ \begin{array}{rrr}
  \Sk{0}{\lambda} & -\Sk{1}{\lambda} &  \Sk{2}{\lambda} \\
  -\Sk{1}{\lambda} & \Sk{2}{\lambda} & -\Sk{3}{\lambda} \\
  \Sk{2}{\lambda}    & -\Sk{3}{\lambda}    & \Sk{4}{\lambda} 
  \end{array}\right]\times \left[ \begin{array}{c} a_2 \\ a_1 \\ a_0 \end{array}\right] $$
or
  $$\bm{m} =  \bm{Z}_{2,\lambda} \times \bm{a}\,. $$ 
We now want to find the value predicted by $P_{2,\lambda,0}$ at time $\tau$. Let $\bm{\tau}_2 = [1,\tau,\tau^2]^T$ be the vector of $\tau$ powers.
Then $u(\tau)$ is equal to $a_2+a_1 \tau +a_0 \tau^2 = \bm{\tau}_2^T\times \bm{a}_2 $. This system might be solved with standard pseudo-inverse techniques~\cite{Greville_T_1960_j-siam-rev_apm}, by premultiplying by the transpose of \Zk{2}{\lambda}. This is not required, 
as \Zk{2}{\lambda} is always invertible, provided that $\lambda > \delta$. Its inverse is denoted \Zk{-2}{\lambda}. It thus does not need to be updated in real-time. It may be computed offline, numerically or even symbolically. Hence:
$$ u(\tau)= \left(\bm{\tau}_2^T \times \Zk{-2}{\lambda}\right) \times \bm{m}_{2,\lambda}\,.$$
The vector $\bm{\tau}_2$ and \Zk{-2}{\lambda} are fixed, and the product $\bm{\tau}_2^T \times \Zk{-2}{\lambda}$ may be stored at once. Thus, for each prediction, the only computations required are  the update of vector $\bm{m}_{2,\lambda}$ and its product with the aforementioned stored matrix.
\subsection{\chopoly symbolic formulation \label{sec:prediction-linear}}
When only one polynomial  predictor is required, actual computations do not require genuine matrix calculus, especially for small degrees $\delta$. With  $\delta = 0$ and  $\omega = 0$ (or $P_{0,\lambda,0}$), one easily sees that:
\begin{equation}
u(\tau)   = \frac{m_0}{\Sk{0}{\lambda}} = \frac{u_0+\cdots+u_{1-\lambda}}{\lambda},\label{eq_0_l_0}
\end{equation}
that is, the running average of past frame values. It reduces to standard ZOH, $u(\tau) = u_0$, when $\lambda = 1$. 
For $\delta = 0$ and  $\omega = 1$, one gets a weighted average giving more importance to the most recent samples:
\begin{equation}
u(\tau) =  2\frac{\lambda u_0 +\cdots+2u_{2-\lambda}+u_{1-\lambda}}{\lambda (\lambda+1)}.\label{eq_0_l_1}
\end{equation}
With  $\lambda = 2$, $\delta = 1$ and  $\omega = 0$, 
$P_{1,2,1}(\tau) = u_{0} + (u_{0} - u_{-1})\tau 
\label{eq_1_2_1}
$ yields the simplest 2-point linear prediction or standard FOH.
$P_{1,3,0}$ yields the simple estimator form: 
\begin{equation}
u(\tau) =   \frac{1}{6}(5u_{0} +2 u_{-1} -u_{-2}) + (u_{0} - u_{-2}) \frac{\tau}{2}\,.
\label{eq_1_3_0}
\end{equation}
For  $P_{2,5,1}$, we tediously get:
{\small
\begin{align}
u(\tau)  = & \frac{1}{70}(65u_{0} + 12u_{-1} - 6u_{-2} - 4u_{-3} + 3u_{-4}) + (25u_{0} - 12u_{-1} - 16u_{-2} - 4u_{-3} + 7u_{-4})\frac{\tau}{28}\nonumber\\ 
&  + (5u_{0} - 4u_{-1} - 4u_{-2} + 3u_{-4})\frac{\tau^2}{28}
\label{eq_2_5_1}
\end{align}}
or with the Ruffini-Horner's method for polynomial evaluation, to slightly reduce the number of operations:
{\small
\begin{align}
u(\tau)  = & \left(\left(\left(25u_{0} - 20u_{-1} - 20u_{-2} + 15u_{-4}\right)\tau+\left(25u_{0} - 12u_{-1} - 16u_{-2} - 4u_{-3} + 7u_{-4}\right)\right)\tau\right.\nonumber\\
&+\left.\left(130u_{0} + 24u_{-1} - 12u_{-2} - 8u_{-3} + 6u_{-4}\right)\right)/140
\,.
\label{eq_2_5_1a}
\end{align}}
One easily remarks that, when the weighting exponent $\omega$ is an integer, prediction polynomials  have rational coefficients, which limits floating-point round-off errors, especially when prediction times and variables (for instance quantized ones)  are integer or rational as well.
\subsection{\chopoly Type I and II computational complexity   \label{sec:typeI-typeII-complexity}}
The computational complexity of a single extrapolation is given in Table \ref{tab_computation}. They are evaluated by a number of elementary operations. They are only meant to provide rough intuitions and guidelines on the actual implemented complexity. Their expressions for Type I  \eqref{eq_typeI} and  Type II  \eqref{eq_typeII} \CHOPO implementations are given terms of  $\delta$, $\lambda$ (and $\omega$). Type I direct implementation is not efficient: unoptimized moments computations yield a cubic complexity in $(\delta,\lambda,\omega)$. Recurring results  can be  evaluated using   call-by-need or lazy evaluation.
For $\mmk{d}{\lambda}{\omega}$, 
the factors $(\lambda - l)^\omega$ yield $\lambda-2$ powers (the  $\lambda -1$ adds are unnecessary), since $0^\omega$ and $1^\omega$ are direct. For $2\le l\le\lambda$, the $l^d$ are gathered in a $(\delta+1) \times \lambda$ array, with $\sum_{d=2}^\delta d = \frac{\delta(\delta+1)}{2}-1$ products.
The  $(\lambda - l)^\omega l^d$ terms can be stored in a $(\delta+1)\times\lambda$ matrix,
involving $(\lambda-2)(\delta-1)$ products. These estimates are stored in the top-half of Table \ref{tab_computation}, above the dashed line.

Upon hopping frame update and $\tau$ determination,  using the simplification $\tau^{d+1} = \tau\tau^{d}$,  $\bm{\tau}_\delta$ requires $\delta -1$ products. The evaluation of the weighted moments entails only $\delta (\lambda-1)$ adds, and $(\lambda-2)\delta+1$ products, since the lazy matrix storing  $(\lambda - l)^\omega l^d$ contains some zeroes and ones (Table \ref{tab_computation}, bottom-half). Both Type I and II are thus roughly quadratic in $(\delta,\lambda)$. Both are competitive with respect to Lagrange extrapolation, which requires $\bigO{\delta^2}$ or $\bigO{\delta}$ with $\delta =\lambda +1$ depending on the implementation.

If we precisely compute  the operation excess from  Type I to II, we obtain $\Xi(\delta,\lambda) = 2\delta^2+\delta+3 -2\lambda$. For the parameters given in \ref{sec:prediction-typeII-examples}, we have for instance $\Xi(0,2)=-1$, $\Xi(1,3)=0$, $\Xi(2,5)=3$. 
\begin{table}[!htb]
\begin{center}
\begin{tabular}{|c|x{1.3cm}|x{1.95cm}|x{1.3cm}|x{1.95cm}|x{1.3cm}|} \hline
Operations & $+$ & $\times$ & power & $+$  &  $\times$  \\  
\hline
 $(\lambda \mathord{-} l)^\omega$ &  &   & $\lambda\mathord{-}2$ &   &   \\  
$l^d$ &  & $\delta(\delta\mathord{-}1)/2$ &   &    &    \\  
$(\lambda \mathord{-} l)^\omega l^d$ &  & $(\lambda\mathord{-}2)(\delta\mathord{-}1)$ &   &    &    \\  
\hdashline
$\bm{\tau}_\delta$ &  & $\delta\mathord{-}1$   & &  & $\delta\mathord{-}1$   \\
$\MMk{\delta}{\lambda}{\omega}$ & $\delta (\lambda\mathord{-}1)$ & $(\lambda\mathord{-}2)\delta\mathord{+}1$ &  &  &  \\
$  \ZZk{-\delta}{\lambda}{\omega} \MMk{\delta}{\lambda}{\omega}$/$\bm{\Pi}_{\delta,\lambda,\omega}\bm{u}_{\lambda} $  & $\delta(\delta\mathord{+}1)$  &  $(\delta\mathord{+}1)^2$ & & $(\delta\mathord{+}1)(\lambda\mathord{-}1)$  &  $(\delta\mathord{+}1)\lambda$   \\
$\bm{\tau}_\delta \times \cdots$ & $\delta$  &  $\delta$ & & $\delta$  &  $\delta$   \\
\hline\hline
Leading orders &  \multicolumn{3}{c|}{Type I: $2(\delta\lambda \mathord{+} \delta^2)$}  &   \multicolumn{2}{c|}{Type II: $2\delta\lambda$} \\
\hline
\end{tabular}
\caption{Elementary  operations required for  $u(\tau)$ in Type I and II implementations. Top: lazy evaluation (computed once). Bottom: required for each hopping frame.\label{tab_computation}}
\end{center}
\end{table}
\subsection{Examples for Type II \chopoly implementation \label{sec:prediction-typeII-examples}}
 Table~\ref{tab_predictors_diff_omega} provides examples of predictor matrices $\bm{\Pi}_{\delta,\lambda,\omega}$ (\emph{cf.} 	\eqref{eq_typeII}) of fixed degree $\delta$ and length $\lambda$, for different integers and rational powers  $\omega \in \left\{0, \frac{1}{8}, \frac{1}{4}, \frac{1}{2},1, 2 \right\}$. 
\begin{table}[!htb]
\begin{center}
\begin{tabular}{cccc}
\toprule
$\omega$ & $\bm{\Pi}_{0,2,\omega}$ & $\bm{\Pi}_{1,3,\omega}$ & $\bm{\Pi}_{2,5,\omega}$\\
\midrule
0 & 
$ \frac{1}{2} \ar\left[ \begin{array}{rr}  1 & 1 \end{array}\right]$ & 
$ \frac{1}{6} \ar\left[ \begin{array}{rrr}  5 & 2 & -1\\ 3&  0 & -3\end{array}\right]$ & 
$ \frac{1}{70} \ar\left[ \begin{array}{rrrrr}  62 & 18 & -6 & -10 & 6\\ 54 & -13  & -40  & -27 & 26\\ 10 & -5  & -10  & -5 & 10\end{array}\right]$  \\
$\frac{1}{8}$ & 
 $\left[ \ar\begin{array}{rr}
  \num{0.5216473} &  \num{0.4783527}
\end{array}\right]$ & 

$\left[ \ar\begin{array}{rrr}
   \num{0.8426476}  & \num{0.3147049} & \num{-0.1573524} \\
   \num{0.5115814}  &\num{-0.0231627} & \num{-0.4884186}
\end{array}\right]$ & 

$\left[ \ar\begin{array}{rrrrr}
   \num{0.8917465}  &\num{ 0.2455946} & \num{-0.0872630}  & \num{-0.1292439} & \num{ 0.0791658}\\
   \num{0.7860286} &\num{-0.2135760} & \num{-0.5754432} & \num{-0.3524997}  & \num{0.3554904}\\
   \num{0.1468234} &\num{-0.0789864} & \num{-0.1439816} & \num{-0.0623714}  & \num{0.1385160}
\end{array}\right]$ \rule[2pt]{0pt}{25pt} \\
$\frac{1}{4}$ & 
$\left[ \ar\begin{array}{rr}
   \num{0.5432136} &  \num{0.4567864}
\end{array}\right]$ &
$\left[ \ar\begin{array}{rrr}
   \num{0.8516937} &  \num{0.29661260} & \num{-0.1483063}\\
   \num{0.5234379} & \num{-0.04687577} & \num{-0.4765621}
\end{array}\right]$ &
$\left[ \ar\begin{array}{rrrrr}
   \num{0.89758532} &  \num{0.2342639} & \num{-0.0883036} & \num{-0.1165257}  & \num{0.0729801}\\
   \num{0.80080398} & \num{-0.2421585} & \num{-0.5783484} & \num{-0.3200437}  & \num{0.3397466}\\
   \num{0.15092258} & \num{-0.0869041} & \num{-0.1448233} & \num{-0.0533315}  & \num{0.1341363}
\end{array}\right]$\rule[2pt]{0pt}{25pt}  \\
$\frac{1}{2}$ & 
 $\left[ \ar\begin{array}{rr}
   \num{0.5857864} &  \num{0.4142136}
\end{array}\right]$ &
$\left[ \ar\begin{array}{rrr}
   \num{0.8689561} &  \num{0.2620878}  &\num{-0.1310439}\\
   \num{0.5479654} & \num{-0.0959308} & \num{-0.4520346}
\end{array}\right]$ &
 $\left[ \ar\begin{array}{rrrrr}
   \num{0.90868344} &  \num{0.2122978} & \num{-0.0889939} & \num{-0.0936393} &  \num{0.0616519}\\
   \num{0.83086766} & \num{-0.3014847} & \num{-0.5807518} & \num{-0.2575129} &  \num{0.3088818}\\
   \num{0.15954032} & \num{-0.1038855} & \num{-0.1455855} & \num{-0.0353338} &  \num{0.1252645}
\end{array}\right]$ \rule[2pt]{0pt}{25pt}  \\
1 & 
$ \frac{1}{3} \ar\left[ \begin{array}{rr} 2  & 1 \end{array}\right]$ & 
$ \frac{1}{10} \ar\left[ \begin{array}{rrr} 9  & 2 & -1\\ 6& -2 & -4\end{array}\right]$ & 
$ \frac{1}{140} \ar\left[ \begin{array}{rrrrr}  130 & 24 & -12 & -8 & 6\\ 125 & -60  & -80  & -20 & 35\\ 25 & -20 & -20 & 0  & 15 \end{array}\right]$\rule[2pt]{0pt}{25pt}  \\
2 & 
$ \frac{1}{5} \left[ \ar\begin{array}{rr} 4 & 1\end{array}\right]$ &
$ \frac{1}{38} \left[ \ar\begin{array}{rrr} 36  & 4 & -2\\ 27& -16 & -11\end{array}\right]$ &
$ \frac{1}{10164} \left[ \ar\begin{array}{rrrrr}  9750 & 1056 & -684 & -144 & 186 \\ 10375 & -7216 &  -5028 & 368 & 1501\\ 2275 & -2464  &  -1176 & 644 & 721 \end{array}\right]$ \rule[2pt]{0pt}{25pt} \\
\bottomrule
\end{tabular}
\caption{Type II matrices  $\bm{\Pi}_{0,2,\omega}$, $\bm{\Pi}_{1,3,\omega}$ and $\bm{\Pi}_{2,5,\omega}$, with  integer and rational weighting powers $\omega \in  
\left\{0, \frac{1}{8}, \frac{1}{4}, \frac{1}{2}, 1, 2 
\right\}$.	\label{tab_predictors_diff_omega}}
\end{center}\end{table}



\clearpage

\bibliographystyle{unsrt}







\end{document}